\documentclass[a4paper,11pt]{article}
\pdfoutput=1 % if your are submitting a pdflatex (i.e. if you have
\usepackage{jcappub} % for details on the use of the package, please
\usepackage[T1]{fontenc} % if needed
\usepackage{ulem}

\usepackage[outline]{contour}
\usepackage{xcolor,graphicx}
\usepackage{caption}
\usepackage{subcaption}

\newcommand{\beq}{\begin{equation}}
\newcommand{\eeq}{\end{equation}} %\indent}
\newcommand{\eei}{\end{equation}\indent\indent}
\newcommand{\bc}{\begin{center}}
\newcommand{\ec}{\end{center}}
\newcommand{\bea}{\begin{eqnarray}}
\newcommand{\eea}{\end{eqnarray}}
\newcommand{\ba}{\begin{array}}
\newcommand{\ea}{\end{array}}

\def\case#1/#2{\textstyle\frac{#1}{#2} }

\newcommand{\be}{\begin{equation}}
\newcommand{\ee}{\end{equation}}

\newcommand{\HH}{\mathcal{H} }

\newcommand{\bn}{{ \bf n }}

\newcommand{\inspire}[1]{\href{http://inspirehep.net/search?p=find+J+#1}
 {{\color{black}[{\color{blue} {\small in}SPIRE}]}}}
\newcommand{\book}[1]{\href{http://inspirehep.net/search?p=#1}
 {{\color{black}[{\color{blue} {\small in}SPIRE}]}}}

\newcommand{\inspired}[1]{\href{http://inspirehep.net/search?p=#1}
 {{\color{black}[{\color{blue} {\small in}SPIRE}]}}}

\contourlength{0.4pt} % adjusts boldness by using \contour{black}{ text }
\contournumber{10} % ... by reprinting it with n-times with an offset

\title{ Lensing convergence 
 in galaxy clustering in \contour{black}{$\Lambda$}CDM and beyond
}

\author[a, b]{Eleonora Villa,}
\author[c, d,e]{Enea Di Dio,}
\author[a, b]{Francesca Lepori,}

\affiliation[a]{SISSA- International School for Advanced Studies, \\ Via Bonomea 265, 34136 Trieste, Italy}
\affiliation[b]{INFN, Sezione di Trieste,\\ Via Valerio 2, I-34127 Trieste, Italy}
\affiliation[c]{INAF - Osservatorio Astronomico di Trieste,\\  Via G. B. Tiepolo 11,  I-34143 Trieste, Italy}
\affiliation[d]{Lawrence Berkeley National Laboratory,\\ 1 Cyclotron Road, Berkeley, CA 93720, USA}
\affiliation[e]{Berkeley Center for Cosmological Physics,\\ University of California, Berkeley, CA 94720, USA}

\emailAdd{evilla@sissa.it}
\emailAdd{enea.didio@berkeley.edu}
\emailAdd{flepori@sissa.it}

\abstract{
We study the impact of neglecting lensing magnification in galaxy clustering analyses for future galaxy surveys, considering the $\Lambda$CDM model and two extensions: massive neutrinos and modifications of General Relativity.
Our study focuses on the biases on the constraints and on the estimation of the cosmological parameters.
We perform a comprehensive investigation of these two effects for the upcoming photometric and spectroscopic galaxy surveys Euclid and SKA for different redshift binning configurations. We also provide a fitting formula for the magnification bias of SKA.
Our results show that the information present in the lensing contribution does improve the constraints on the modified gravity parameters whereas the lensing constraining power is negligible for the
$\Lambda$CDM parameters.
For photometric surveys the estimation is biased for all the parameters if lensing is not taken into account. This effect is particularly significant for the modified gravity parameters. Conversely for spectroscopic surveys the bias is below one sigma for all the parameters. 
Our findings show the importance of including lensing in galaxy clustering analyses for testing General Relativity and to constrain the parameters which describe its modifications.
}

\begin{document}
\maketitle
\flushbottom

%%%%%%%%%%%%%%%%%%%%%%%%%%%%%%%%%%%%%%%%%%%%%%%%%%%%%%%%%%%%%%%%%%%%%%%
%%%%%%%%%%%%%%%%%%%%%%%%%%%%%%%%%%%%%%%%%%%%%%%%%%%%%%%%%%%%%%%%%%%%%%%
\section{Introduction}
\label{sec:into}
%%%%%%%%%%%%%%%%%%%%%%%%%%%%%%%%%%%%%%%%%%%%%%%%%%%%%%%%%%%%%%%%%%%%%%%
%%%%%%%%%%%%%%%%%%%%%%%%%%%%%%%%%%%%%%%%%%%%%%%%%%%%%%%%%%%%%%%%%%%%%%%%
Cosmology is entering the so-called precision era, where cosmological parameters are aimed to be constrained at the  percent level. While current constraints are mainly driven by CMB experiments, in particular Planck~\cite{Ade:2015xua, Ade:2015rim}, in the next decade different Large Scale Structures (LSS) surveys will map almost the whole observable universe at low redshift. Beyond constraining the standard $\Lambda$CDM parameters with an unprecedented accuracy, we have the opportunity to test the theory of gravity at large scales.
The predictions of General Relativity (GR) have been successfully tested with great precision in the near Universe, in the weak-field regime of
the solar system~\cite{Everitt:2011hp, Everitt:2015qri} and in the strong-field regime of binary pulsar systems~\cite{Taylor:1989sw, Will:2014kxa}, black hole~\cite{Abbott:2016blz, TheLIGOScientific:2016src, Abbott:2017vtc} and neutron star~\cite{TheLIGOScientific:2017qsa} mergers. Cosmological observations allow GR to be investigated on entirely new length and time scales. Testing GR on cosmological scales is further motivated by the lack of a convincing physical explanation for the accelerated expansion and by the purpose of comparing data with Dark Energy or modified gravity models.
Moreover, several modified gravity models give rise naturally to some screening mechanisms, which due to non-linearity freeze the additional degree(s) of freedom on small scales recovering the GR predictions. This provides a further motivation to test GR on cosmological scales, i.e.~well beyond the screening radius, where it is possible to discriminate between GR and different modified gravity models.

The success of precision cosmology depends not only on precise observations, but also on the theoretical modelling, which must be understood to at least to the same level of accuracy to avoid theoretical systematics that may bias the measurements at a level that cannot be neglected nowadays. 
In this work we focus on a key cosmological observable, namely the galaxy number counts.
It receives contributions from the density perturbation, 
from the peculiar velocity of the sources, the well-known redshift space distortions and the Doppler term \cite{kaiser}, and from the fact that the photons travel on perturbed geodesics, leading to perturbations in the observed redshift and in the spacetime volume.
These additional contributions, beyond density perturbations and redshift space distortions, are the so-called relativistic effects and they are due to gravitational lensing, Doppler, Shapiro time-delay and integrated Sachs-Wolfe effects on the path of photons. See Refs.~\cite{Yoo:2009,Yoo:2010,Bonvin:2011bg,Challinor:2011bk,Schmidt:2012ne, Bertacca:2012, Raccanelli:2013gja, Raccanelli:2013} for the fully relativistic computation at linear order in the perturbations.
In our analysis we only consider the lensing magnification effect, which enters at the same parametrical order (in terms of an expansion in $\HH /k$) of the standard Newtonian contributions: density and redshift space distortion. It accounts for two competing modifications to the galaxy number counts due to gravitational lensing. At one side the number of observed galaxies is decreased because of the stretching of the solid angle of observation. At the other side, it can be increased (or decreased) since the observed luminosity is affected by matter perturbations in a way that the signal of a galaxy behind overdense regions is magnified (or de-magnified for galaxies behind underdense regions) and therefore can be detected, even if the intrinsic luminosity is below the threshold of the survey.
The last effect, which takes into account that a real galaxy survey is luminosity-limited, is called magnification bias and depends on the specifications of the survey. The lensing magnification effect has been predicted more than two decades ago~\cite{Bartelmann:1994ye,Dolag:1997yt,Sanz:1997pw} and first detected in 2005~\cite{Scranton:2005ci}. Whereas other relativistic effects has been shown to be detectable by cross-correlating different probes~\cite{mcdonald:2009,Yoo:2012se, Bonvin:2013, Alonso:2015sfa,Fonseca:2015laa, Irsic:2015, Bonvin:2015,Gaztanaga:2015, Dai:2015wla, Borzyszkowski:2017ayl, Abramo:2017xnp} (see however~\cite{Raccanelli:2016avd}), their impact on a single tracer analysis can be safely neglected, apart for the measurement of primordial non-Gaussianity (see~\cite{Camera:2014bwa, Camera:2014sba, Raccanelli2015,Alonso:2015uua}) that we do not address.

While lensing magnification effect has been known by decades, ongoing galaxy clustering surveys are not including its contribution in the data analysis process. In view of the next generation of surveys which will reach higher redshifts, we are interested in questioning the validity of this approximation.
This issue has been considered already in Refs.~\cite{Camera:2014bwa, Camera:2014sba, Raccanelli2015, Montanari:2015rga, DiDio:2016ykq, Cardona:2016qxn,Lepori:2016rzi}. 
The main result is that including the lensing contribution to the galaxy clustering analysis is essential in order to avoid biased estimations of the best-fit values and constraints, especially for some cosmological parameters. In our work we focus on the impact on parameters beyond $\Lambda$CDM: massive neutrinos and modified gravity.
To test GR, and consequently to constrain the cosmological parameters which describe the deviation from GR, observations need to measure both the Bardeen potentials. While the standard Newtonian contribution to galaxy clustering is sensitive only the gravitational potential described by the time component of the metric, which determines the motion of non-relativistic object as galaxies, the lensing magnification measures the Weyl potential, hence the sum of the two Bardeen potentials. Therefore, it is clear that the lensing magnification carries the useful information to test for deviation of GR and it is worth to investigate if neglecting its contribution to the galaxy number counts may bias the results. This issue was addressed in Ref.~\cite{Alonso:2016suf}, where the authors study the impact of lensing and the other relativistic effects on the constraints for the case of a sub-class of Horndeski scalar-tensor theories, also in combination with other probes.

In our work we will consider the parametrized description where modifications of gravity are encoded in two parameters that can be in general functions of time and scale: they modify the Poisson equation and the relation between the two Bardeen potentials prescribed by GR. We will study the impact of neglecting lensing magnification in the constraints and also in the best-fit value of the cosmological and modified gravity parameters inferred from galaxy clustering measurements. We will also analyse the dependence of the lensing contribution on the type of survey and on the redshift binning configuration. Other works~\cite{Raccanelli2015, Montanari:2015rga, DiDio:2016ykq, Cardona:2016qxn} have described the lensing magnification impact in terms of few very broad redshift bins, which can adapt well to photometric surveys. We will extend this analysis to modified gravity. Moreover we will study how the lensing impact changes by increasing the number of redshift bins. While we can not reach the same redshift accuracy of a spectroscopic survey, we intend to find a hint of the real effect on spectroscopic surveys. In this work we choose to use the photometric specifications for Euclid\footnote{ \url{http://sci.esa.int/euclid/}} and the spectroscopic specifications for SKA\footnote{\url{http://skatelescope.org/}}.

The paper is organized as follows. We introduce the galaxy number counts and describe the parametrization we use for modified gravity in section~\ref{sec:GNC} and describe the methodology and the survey specifications that we use in section~\ref{sec:method}. We then present our results in section~\ref{sec:results} and draw our conclusions in section~\ref{sec:conclusions}. In Appendix~\ref{sec:table} we report explicitly some of our results.

%%%%%%%%%%%%%%%%%%%%
\section{Galaxy Number Counts}
\label{sec:GNC}

In a galaxy clustering experiment cosmologists measure the number of galaxies or sources $N( \bn, z )$ in terms of the observed direction $\bn$ and measured redshift $z$. We can therefore naturally define the galaxy number count observable as
\be
\Delta_{\rm gal} (\bn , z) = \frac{ N ( \bn , z) - \langle N \rangle (z) }{ \langle N \rangle (z)}
\ee
where $\langle .. \rangle$ denotes the angular average at fixed measured redshift $z$. The galaxy number counts at first order in perturbation theory reads
\bea \label{number_counts}
\Delta_{\text{gal}} (\mathbf{n}, z) &=& b_\text{gal} D  + \frac{1}{\mathcal{H}(z)} \partial_r (\mathbf{V}\cdot\mathbf{n}) \notag
\label{nc_gal}\\
&&  + (5s - 2) \int_0^{r(z)} \frac{r(z) - r}{2 r(z) r} \Delta_{\Omega} (\Phi + \Psi) dr \notag \\
&&+ \Biggl(\frac{\mathcal{H'}}{\mathcal{H}^2} + \frac{2-5 s}{r\mathcal{H}} + 5 s - f^{\text{gal}}_{\text{evo}}
\Biggr)(\mathbf{V}\cdot \mathbf{n} ) \notag \\
&&+( f^{\text{gal}}_{\text{evo}}  - 3)\HH V +(5s - 2) \Phi + \Psi + \frac{1}{\mathcal{H}} \Phi' +\frac{2-5s}{r(z)} \int^{r(z)}_0 dr (\Phi + \Psi) \notag \\
&& +\Biggl(\frac{\mathcal{H'}}{\mathcal{H}^2} + \frac{2-5 s }{r(z)\mathcal{H}}+5s  - f^{\text{gal}}_{\text{evo}}\Biggr)\Biggl(\Psi + \int^{r(z)}_0 dr (\Phi' + \Psi')\Biggr) \, ,       
\eea
where $D$ is the dark matter density fluctuation in synchronous gauge, ${\bf V}$ the peculiar velocity in longitudinal gauge, $V$ the potential velocity related through ${\bf V} = - \nabla V$ to the peculiar velocity and $\Phi$ and $\Psi$ are the gauge-invariant Bardeen potentials. A prime indicates the partial time derivative with respect to conformal time $\eta$ and $\HH= a'/a$ is the conformal Hubble parameter. The Eq.~\eqref{number_counts} introduces three bias parameters to relate dark matter perturbations to galaxies, namely a galaxy bias $b_{\rm gal}$, a magnification bias defined as
\begin{equation} \label{magnbias}
s(z, F_*)= -\frac{2}{5} \frac{\partial\, {\rm ln}\,\bar{n} (z, F > F_*)}{\partial\, {\rm ln}F_*}
\end{equation}
where $\bar{n}$ is the background cumulative number density of galaxies with flux above the threshold $F_*$, namely
\begin{equation} \label{nflux}
\bar{n} (z, F > F_*)= \int^\infty_{{\rm ln}F_*} \bar{n} (z, {\rm ln}F)\, d\,{\rm ln}F\,,
\end{equation}
and an evolution bias 
\be
f_{\rm evo}^{\rm gal}\left( z, F>F_*\right) = \frac{\partial \ln \bar n  \left( z, F> F_* \right) }{\HH \partial \eta}.
\ee

Being the galaxy number counts a function of a unit vector $\bn$ and a redshift $z$, it is natural to expand them in terms of spherical harmonics for different redshifts
\be
\Delta_{\rm gal} (\bn , z)  = \sum_{\ell m} a_{\ell m} \left( z \right) Y_{\ell m} \left( \bn \right)
\ee
with 
\be
a_{\ell m} \left( z \right) = \int d\Omega_{\bn} Y^*_{\ell m } \left( \bn \right) \Delta_{\rm gal} (\bn , z) 
\ee
This leads to the redshift-dependent galaxy power spectra
\be
\langle a_{\ell m}\left( z \right) a^*_{\ell' m'} \left( z'\right) \rangle = C_\ell\left( z, z' \right) \delta_{\ell \ell'} \delta_{m m'}
\ee
where the Kronecker symbols are a consequence of angular isotropy. It is worth remarking that, despite working with angular statistics, the power spectra $C_\ell\left( z, z' \right)$ contain the full 3-dimensional information, which can be fully recovered by optimally cross-correlating spectra at different redshifts, see e.g.~Refs.~\cite{Asorey:2012rd,DiDio:2013sea}.
In order to cross-correlate different redshifts we need to introduce a redshift binning, hence we defined the binned spectra as
\be \label{eq:clij}
C^{ij}_\ell = \int dz'_1 dz'_2 \frac{d N}{dz'_1}  \frac{d N}{dz'_2} W_i \left( z'_1; z_1 , \sigma_1 \right)  W_j\left( z'_2; z_2 , \sigma_2 \right) C_\ell \left( z'_1, z'_2 \right)
\ee
where $dN/dz$ is the survey selection function and $W_i$ is the window function for the $i-$th redshift bin centred at $z_i$ with a width of $\sigma_i$ normalised to the unity. In our work we will consider two different shapes of window functions: tophat and gaussian bins, see section~\ref{sec:surveys}.

In our analysis we consider the first three terms of the galaxy number counts in Eq.~\eqref{number_counts}: beside the contribution from the density, we include only redshift-space distortions and lensing which physically represent the leading perturbations of the space-time volume, in the radial and transverse direction, respectively.
The lensing magnification effect enters at the same parametrical order (in terms of an expansion in $\HH /k$) of the standard Newtonian contributions - density and redshift-space distortions - and is expected to dominate over the other relativistic effects for typical current and upcoming galaxy surveys. 
It is well known that it dominates radial cross-correlations between different redshift bins, while the contributions due to the Newtonian terms drops rapidly.
The lensing effect can be described easily in terms of angular power spectra, as done for CMB or shear weak lensing, but including lensing in a 3-dimensional Fourier analysis is a complicate and severe problem~\cite{Mandel:2005xh}. Indeed lensing gets contributions along the whole line of sight and therefore it mixes non-trivially different Fourier scales. 
As we see from Eq.~\eqref{number_counts}, lensing magnification is the angular gradient of the lensing potential to the source position. Hence the lensing magnification is a pure transverse effect, see Ref.~\cite{DiDio:2016kyh} for the impact of lensing on the transversal modes of the matter power spectrum, and by adding more redshift correlations we do not measure more physical modes induced by lensing. We therefore expect that by increasing the number of redshift bins, or having a better redshift resolution, will not increase the number of modes induced by lensing but only the ones dominated by density and redshift space distortions. This simple picture neglect the information in the redshift evolution, but to first approximation we do expect that the relative impact of lensing magnification decreases by measuring more radial modes, apart for the cosmological parameters which require the lensing information.
In the next sessions we perform a Fisher analysis to provide a quantitative answer.

We compute the angular power spectra by using a modified version\footnote{\url{https://gitlab.com/philbull/mgclass}} of {\sc class}~\cite{Lesgourgues:2011re,CLASSgal,Baker:2015bva} in order to include modifications of GR. We consider first-order scalar metric perturbations about a flat Friedmann-Lema\^itre-Robertson-Walker (FLRW)
background. The line element in the Newtonian gauge reads
\begin{equation}
{\rm ds}^2 = a^2 (\eta) \left[ -\left(1+2\Psi \right)  {\rm d}\eta^2 +(1-2\Phi)\delta_{ij}{\rm d}x^i{\rm d}x^j  \right]\,,
\end{equation}
where $\eta$ is the conformal time, $a$ is the scale factor, and $\Psi$ and $\Phi$ are the Bardeen potentials. 
In GR the linearized Einstein equations  set the two Bardeen potentials $\Psi$ and $\Phi$ to be equal (in absence of anisotropic stress) and relate them to the matter density in comoving gauge through the Poisson equation.
Modified theories of gravity can change  these relations. It is therefore common to parametrize deviations from GR by the following relations~\cite{Zhao:2008bn}:
\begin{align}
\frac{\Phi}{\Psi}&= \gamma  \label{slip}\\
k^2 \Psi &= - \frac{3}{2}\frac{\mathcal{H}_0^2\Omega_{m_0}}{a}\mu  D
\,, \label{mu}
\end{align}
where $\Omega_{m_0}$ and $\mathcal{H}_0$ are the density and Hubble parameter today, respectively. 
The ``clustering parameter'' $\mu$ parametrizes modification of the Poisson equation relating the gauge invariant density perturbation $D$ to the potential $\Psi$ and the ``slip parameter'' $\gamma$ parametrizes anisotropic stress which makes the ratio between the two scalar potentials different from unity. In this paper we focus on the effects of lensing magnification which depends the Weyl potential $\Phi + \Psi$. Thus we will consider the equivalent parametrization in terms of $\left\{\mu, \Sigma\right\}$ instead of the one in $\left\{\gamma, \mu\right\}$, where $\Sigma$ is defined by
\begin{equation}
\Sigma = - \frac{ k^2 \left(\Phi + \Psi\right)}{3\mathcal{H}_0^2\Omega_{m_0} D} = \frac{\mu\left(1+\gamma\right)}{2}\,. \label{sigma}
\end{equation}
In GR $\mu=\gamma=\Sigma=1$ whereas in modified gravity theories, in full generality, they can depend on time and wave number. Throughout this work we consider the most simple case, where $\mu$ and $\Sigma$ assume constant values $\mu_0$ and $\Sigma_0$ at late times, starting from $\mu=\Sigma=1$ at early times. We fix the threshold redshift at which gravity starts to be modified from GR at $z = 5$.

%%%%%%%%%%%%%%%%%%%%%%%%%%%%%%%%%%%%%%%%%%%%%%%%%%%%%%%%%%%%%%%%%%%%%%%
%%%%%%%%%%%%%%%%%%%%%%%%%%%%%%%%%%%%%%%%%%%%%%%%%%%%%%%%%%%%%%%%%%%%%%%
\section{Methodology}
\label{sec:method}
%%%%%%%%%%%%%%%%%%%%%%%%%%%%%%%%%%%%%%%%%%%%%%%%%%%%%%%%%%%%%%%%%%%%%%%
%%%%%%%%%%%%%%%%%%%%%%%%%%%%%%%%%%%%%%%%%%%%%%%%%%%%%%%%%%%%%%%%%%%%%%%
In order to estimate the bias on the constraints 
and on the best fit 
of the cosmological parameters we use the Fisher matrix formalism, see \cite{Fisher:1935bi,Tegmark:1997yq}. 
The Fisher matrix for the galaxy angular power spectra is given by
\begin{equation} \label{Fishdef}
F_{\alpha\beta}=  \sum^{\ell _{\rm max}}_{\ell={\ell _{\rm min}}} \sum_{(ij)(pq)} \frac{\partial C^{ij}_{\ell,{\rm th}}}{\partial\Phi_\alpha}  \frac{\partial C^{pq}_{\ell, {\rm th}}}{\partial\Phi_\beta} \sigma^{-2}_{C_{\ell}\,[(ij),(pq)]}\,,
\end{equation}
where $\Phi_\alpha$ denotes the $\alpha$-th parameter, the second sum over runs over the indices $(ij)$ and $(pq)$ with $i \leq j$ and  with $p \leq q$ which range from $1$ to the total number of redshift bins $N_{\rm bin}$, and all bin auto- and cross-correlations are included, unless stated differently.
For Gaussian fluctuations the error in the measured power spectra is given by, \cite{DiDio:2013sea}
\begin{equation}
\sigma^{2}_{C_{\ell}\,[(ij),(pq)]}= \frac{C^{(ip)}_{\ell, {\rm obs}} C^{(jq)}_{\ell, {\rm obs}} +  C^{(iq)}_{\ell, {\rm obs}} C^{(jp)}_{\ell, {\rm obs}}  }{(2\ell +1)f_{\rm sky}}\,.
\end{equation}
The observed correlation multipoles  $C^{ij}_{\ell, {\rm obs}}$ including shot-noise are
\begin{equation}
C^{ij}_{\ell, {\rm obs}} = C^{{\rm Nwt}}_{\ell} + C^{{\rm Lens}}_{\ell} + \delta^{ij} N_i
\end{equation}
where $N_i = d N_{z_i}/d\Omega$ denotes the number of sources per steradian in the $i$-th bin. 
We remark that we always include the lensing convergence contribution in the observed power spectra in the covariance. We choose $\ell _{\rm min}=2$ and $\ell _{\rm max}=300$,
and we comment on the dependence on $\ell_{\rm max}$ in Section~\ref{sec:Rshifts}.
It is worth remarking that more sophisticate redshift dependent non-linear cutoffs can be chosen, see e.g.~\cite{DiDio:2013sea,Alonso:2016suf}. Nevertheless our choice, namely to use the same $\ell_{\rm max}$ for any redshift-bin is conservative. Indeed this includes a larger amount of small transversal scales at low redshift compare to high redshift. Therefore it includes more modes in the regime where the lensing magnification is less relevant.

Regarding the theoretical power spectra we will instead assume two different models. As in \cite{Alonso:2015uua,DiDio:2016ykq} we introduce an additional parameter $\epsilon_L$ and write the power spectra as
\begin{align}\label{defeL}
C_{\ell,{\rm th}}(\theta_\alpha, \epsilon_L)  
&= C_{\ell}^{\rm g}(\theta_\alpha) +C_{\ell}^{\rm RSD}(\theta_\alpha) + \epsilon_L C_{\ell}^{\rm Lens} (\theta_\alpha) \nonumber \\
&= C_{\ell}^{\rm Nwt}(\theta_\alpha) +\epsilon_L C_{\ell}^{\rm Lens} (\theta_\alpha)  
\end{align}
where the first two terms represent galaxy clustering and redshift-space distortions and
we have explicitly shown the dependence on the cosmological parameters $\{\theta_\alpha\}$ and on the lensing parameter $\epsilon_L$.
When the galaxy number counts are modelled by the Newtonian terms only, i.e.~$\epsilon_L=0$, we just consider the first two terms in Eq.~\eqref{defeL}.
When instead we include the lensing contribution $\epsilon_L=1$ and the power spectra are given by
\begin{align} \label{ClthLens}
C_{\ell, {\rm th}}(\theta_\alpha) 
&= C_{\ell}^{\rm Nwt}(\theta_\alpha) + C_{\ell}^{\rm Lens} (\theta_\alpha) \,.
\end{align}
For convenience we split the parameter set as $\{\Phi\}=\{\theta_\alpha,\epsilon_L \}$. The Fisher matrix $F$ for the full set of parameters $\{\theta_\alpha,\epsilon_L\}$ has then the form
\begin{equation} \label{Fisherfull}
F^{\Phi\Phi} = \begin{pmatrix}
 F^{\theta\theta}&  F^{\theta\epsilon_L} \cr
 F^{\theta\epsilon_L}& F^{\epsilon_L\epsilon_L} 
 \end{pmatrix}\,.
\end{equation}
The derivatives with respect to the cosmological parameters $\{\theta_\alpha\}$ are approximated using a five-point stencil, 
where for the step size $\delta\theta_\alpha$ we take the $10\%$ of the fiducial value and we have verified that our results do not depend on the particular choice for $\delta\theta_\alpha$.
The derivatives with respect to the lensing parameter $\epsilon_L$ are calculated analytically and are simply given by $C_\ell^{\rm Lens} $. 

We assume a fiducial flat $\Lambda$CDM consistent with Planck, \cite{Ade:2015xua}, including massive neutrinos. The fiducial values of the cosmological parameters are as follows. The amplitude of the curvature fluctuations is ${\rm ln}10^{10}A_{s}=3.094$ and the spectral index is $n_s=0.9645$. The matter content of the Universe is parametrized by $h^2 \Omega_b=\omega_b=2.225 \times 10^{-2}$, $h^2 \Omega_m=\omega_m=0.1198$, $M_\nu=0.06 \,{\rm eV}$ for the baryon and cold dark matter density parameters and for the sum of the neutrino masses, respectively. The dimension-less Hubble parameter is set to be $h=0.6727$. For the modified gravity parameters the fiducial values are $\mu=\Sigma=1$.
We also consider the galaxy bias at redshift $z=0$, $b_0$, as an additional parameter with fiducial values consistent with the specifications for Euclid and SKA that we assume, see eqs.~\eqref{b0EU} and~\eqref{b0SKA}. We will marginalize all our results with respect to the galaxy bias parameter.
We do not impose any prior from other experiments nor combine galaxy clustering with other probes: our purpose is to focus on a galaxy clustering analysis only and to study the relative impact of the lensing contribution in the determination of the cosmological parameters and in the constraining power of the surveys under consideration.
Shear weak lensing surveys and lensing magnification will provide the same information about the lensing potential, but the two measurements suffer by different systematic effects: on one hand we have intrinsic alignment and on there other one galaxy and magnification biases.

%%%%%%%%%%%%%%%%%%%%%%%%%%%%%%%%%%%%%%%%%%%%%%%%%%%%%%%%%%%%%%%%%%%%%%%
\subsection{Bias on the parameter constraints}
%%%%%%%%%%%%%%%%%%%%%%%%%%%%%%%%%%%%%%%%%%%%%%%%%%%%%%%%%%%%%%%%%%%%%%%
\label{sec:Mconstraints}
Our aim is to quantify the amount of information that is misinterpreted if the lensing contribution is neglected in the galaxy clustering analysis. We have therefore introduced in Eq.~\eqref{defeL} the lensing parameter $\epsilon_L$ but for our purpose here we do not treat it as the other cosmological parameters. 
This is because in our approach $\epsilon_L$ is not free to vary over any value and it is not supposed to be measured by the experiments. Rather we treat the lensing contribution in the galaxy power spectra as a theoretical systematic meaning that the parameter $\epsilon_L$ is fixed to $1$ or to $0$, depending on which is the model we assume with or without the lensing term respectively.
We consider the sub-matrix $F^{\theta\theta}$ in Eq.~\eqref{Fisherfull} for the cosmological parameters only
\begin{equation} \label{cosmofish}
F^{\theta\theta}_{\alpha\beta}=  \sum^{ \ell_{\rm max}}_{\ell=2} \sum _{(ij)(pq)} \frac{\partial C^{ij}_{\ell,{\rm th}}}{\partial\theta_\alpha}  \frac{\partial C^{pq}_{\ell, {\rm th}}}{\partial\theta_\beta}  \sigma^{-2}_{C_{\ell}\,[(ij),(pq)]}\,,
\end{equation}
and estimate the standard marginalized errors for the cosmological parameters from
\begin{equation}\label{errordef}
\sigma_{\theta_\alpha}= \sqrt{\left[\left( F^{\theta\theta}  \right)^{-1}\right]_{\alpha\alpha}}
\end{equation}
We then compare the marginalized errors that we get from \eqref{errordef} with the spectra given by 
$C_{\ell, {\rm th}}(\theta_\alpha) = C_{\ell}^{\rm Nwt}(\theta_\alpha) + C_{\ell}^{\rm Lens} (\theta_\alpha)$ with the ones that we get with the Newtonian spectra only. 
In both the analyses we include the lensing contribution in the observed power spectra in the covariance. The change in the marginalized errors is due to the actual dependence of the lensing term on the cosmological parameters.

The Fisher matrix \eqref{cosmofish} has dimension $N\times (N+1)/2$, where $N$ is the number of cosmological parameters under consideration: we have $5$ parameters for $\Lambda$CDM, $6$ parameters when we include massive neutrinos and $8$ parameters including also modified gravity, parametrized by $\mu$ and $\Sigma$.  The errors in Eq.~\eqref{errordef} on each parameter are calculated marginalizing over all the others.

%%%%%%%%%%%%%%%%%%%%%%%%%%%%%%%%%%%%%%%%%%%%%%%%%%%%%%%%%%%%%%%%%%%%%%%
\subsection{Bias on the parameter estimation}
%%%%%%%%%%%%%%%%%%%%%%%%%%%%%%%%%%%%%%%%%%%%%%%%%%%%%%%%%%%%%%%%%%%%%%%
\label{sec:Mshift}
Our approach for the calculation of the bias on the estimation of the cosmological parameters is based on the Fisher matrix method applied to the case of nested models as introduced in \cite{Taylor:2006aw,Knox:1998fp,Heavens:2007ka,Kitching:2008eq}. Suppose to have two models such that the parameter space of one model is contained in the one of the other model, in a way that both the two models have some parameters in common, e.g.~the cosmological parameters $\{\theta_\alpha\}$, that are supposed to be measured from some experiment. The difference between the two models is that in the ``wrong'' model some other parameters are fixed to some  known values whereas in the ``correct'' model they are free to vary as the other parameters. 
This framework can be adapted to the situation where the additional parameters represent the amplitude of some systematic effects, i.e.~they are set to zero in the wrong model, where systematics are not taken into account or they are set to 1, which corresponds to the correct model, where the effect of systematics is included. This is the case we are studying here: under the general assumption that the Universe is described by the power spectra including the lensing contribution, in the correct model we set $\epsilon_L=1$ whereas in the wrong Newtonian model we have $\epsilon_L=0$.

The wrong model assumption leads to a shift in the best-fit values of the cosmological parameters $\{\theta_\alpha\}$ which follows simply from the fact that the dependence of the likelihood on the additional parameters in the correct model displaces the value of the maximum, \cite{Taylor:2006aw}.
The shift in the best-fit value of a parameter is given by
\begin{equation}\label{shiftformula}
\Delta\theta_\alpha= \sum_{\beta} \left(F^{\theta\theta} \right)^{-1}_{\alpha\beta} F^{\theta\epsilon_L}_{\beta}\,,
\end{equation}
where both the terms are subset of the Fisher matrix $F^{\Phi\Phi}$ of the full set of the parameters of Eq.~\eqref{Fisherfull}: $\left(F^{\theta\theta} \right)^{-1}_{\alpha\beta}$ is the inverse of the sub-matrix of the $\{\theta_\alpha\}$ parameters and $F^{\theta\epsilon_L}$ is the column corresponding to the $\{\theta_\alpha\}$ and $\epsilon_L$ parameters.
The estimation of Eq.~\eqref{shiftformula} is based on a first-order Taylor expansion of the likelihood of the correct model and relies on two crucial assumptions. Firstly the difference between the two models has to be small, and secondly the approximated expression in \eqref{shiftformula} holds locally, i.e.~for small shifts $\Delta\theta_\alpha$: for large shifts the Taylor approximation is not accurate meaning that the results can only be trusted qualitatively.
Keeping in mind this caveat, it is worth to discuss two more subtle issues related to the terms appearing in the Fisher matrix in Eq.~\eqref{shiftformula}. The first concerns which spectra have to be used for the derivatives with respect to the cosmological parameters $\{\theta_\alpha\}$, where we have two possibilities: the spectra with lensing $C_{\ell, {\rm th}}(\theta_\alpha) = C_{\ell}^{\rm Nwt}(\theta_\alpha) + C_{\ell}^{\rm Lens} (\theta_\alpha)$ - fiducial model  $\epsilon_L=1$ - or just the Newtonian ones - fiducial model  $\epsilon_L=0$, where in practice the second choice neglects the fact that the lensing term depends on the cosmological parameters. If we use the spectra with lensing we are calculating the bias in the parameter estimation due to neglecting a systematic effect that depends itself on such parameters. This is indeed the case for lensing and actually we have exploited this dependence to estimate the bias in the constraints, see the discussion of our method in section~\ref{sec:Mconstraints}. However this approach would of course lead to larger shifts $\Delta\theta_\alpha$: we have calculated that in this case the shifts normalized to their fiducial values are much greater than one and roughly one order of magnitude larger than the results using Newtonian spectra for almost all the cosmological parameters and for both Euclid and SKA. Since, as remarked above, the approximated formula \eqref{shiftformula} is accurate for small shifts, we prefer to be more conservative and use the Newtonian spectra for the derivatives with respect to the parameters $\{\theta_\alpha\}$. The second issue is related to the errors $\sigma^{2}_{C_{\ell}}$. Again we have two choices: to include the lensing contribution in the measured spectra or not. In this case we do take into account the lensing contribution simply because, even if it is not included in the theoretical description of the galaxy clustering, it is nevertheless observed. 

To summarize, for the terms in Eq.~\eqref{shiftformula} we take 
\begin{equation} \label{fishmix1}
F^{\theta\theta}_{\alpha\beta}=  \sum^{ \ell_{\rm max}}_{\ell=2} \sum_{(ij)(pq)} \frac{\partial C_{\ell}^{{\rm Nwt}\,ij}}{\partial\theta_\alpha}  \frac{\partial C_{\ell}^{{\rm Nwt}\,pq}}{\partial\theta_\alpha} \sigma^{-2}_{C_{\ell}\,[(ij),(pq)]}\,,
\end{equation}
and
\begin{equation} \label{fishmix2}
F^{\theta\epsilon}_{\alpha\beta}=  \sum^{ \ell_{\rm max}}_{\ell=2} \sum_{(ij)(pq)} \frac{\partial C_{\ell}^{{\rm Nwt}\,ij}}{\partial\theta_\alpha}  C_{\ell}^{{\rm Lens}\,pq} \sigma^{-2}_{C_{\ell}\,[(ij),(pq)]}\,,
\end{equation}
where we remind the reader that $C_{\ell}^{{\rm Lens}}$ is the derivative of the spectra including lensing with respect to $\epsilon_L$ and we include the lensing contribution in the errors  $\sigma^{2}_{C_{\ell}}$. Our choices follow the approach of \cite{Kitching:2008eq} and are slightly different to the one used in previous literature for the same effect, i.e.~neglecting lensing (and the other relativistic) effects, see \cite{Camera:2014sba,DiDio:2016ykq,Cardona:2016qxn}: in these works the lensing contribution was neglected in the modelled (as we do) but also in the observed power spectra.

%%%%%%%%%%%%%%%%%%%%%%%%%%%%%%%%%%%%%%%%%%%%%%%%%%%%%%%%%%%%%%%%%%%%%%%
\subsection{Surveys specifications}
\label{sec:surveys}
\begin{figure}[h!]
%\begin{figure}[!tbp]
  \begin{subfigure}[h!]{0.48\textwidth}
    \includegraphics[width=\textwidth]{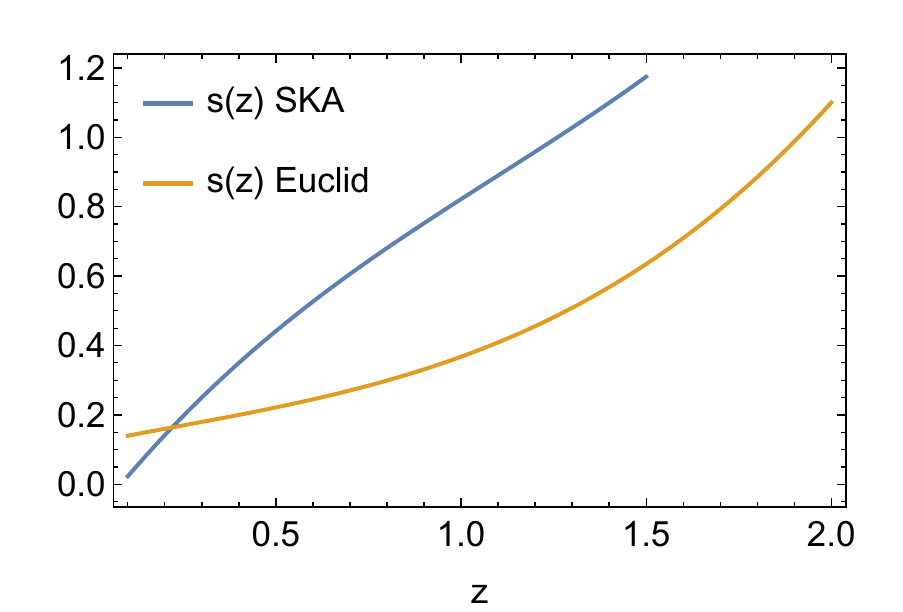}
    %\caption{}
    \label{fig:biasesEU}
  \end{subfigure}
  \hfill
  \begin{subfigure}[h!]{0.48\textwidth}
    \includegraphics[width=\textwidth]{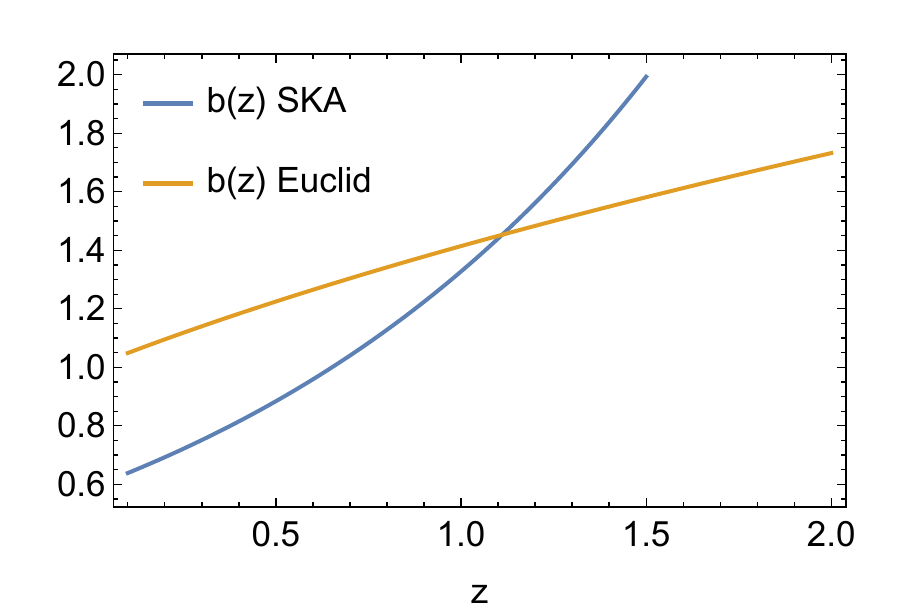}
  %  \caption{}
    \label{fig:biasesSKA}
  \end{subfigure}
  \caption{
  Magnification bias (left panel) and galaxy bias (right panel) for the EUCLID and SKA specification considered in this paper. A constant galaxy and magnification bias is assumed in each redshift bin, with value evaluated at the mean redshift.}
  \label{fig:specifications}
\end{figure}

%%%%%%%%%%%%%%%%%%%%%%%%%%%%%%%%%%%%%%%%%%%%%%%%%%%%%%%%%%%%%%%%%%%%%%%
In this section we present the specifications that we assume for the galaxy surveys  we consider: Euclid and SKA. They are summarized in Fig.~\ref{fig:specifications}.

We consider Euclid photometric specifications as presented in \cite{Amendola:2012ys}: the number of galaxies per redshift and per steradian, the galaxy bias and the magnification bias are given by
\begin{align}
\frac{{\rm d}N}{{\rm d}z{d}\Omega}&=3.5 \times 10^8 z^2 {\rm exp}\left[-\left(\frac{z}{z_0}\right)^{3/2}\right] \\
\text{for}\quad & 0.1 < z < 2 \\
f_{\rm sky}&=0.375 \\
b(z)&=b_0 \sqrt{1+z}\,, \label{b0EU}
\end{align} 
where $z_0=z_{\rm mean}/1.412$, the median redshift is $z_{\rm mean}=0.9$ and we set $b_0=1$. 
We assume the magnification bias as computed in \cite{Montanari:2015rga}:
\begin{equation}
s(z)= s_0 +s_1z+s_2z^2+s_3z^3\,,
\end{equation}
where the coefficients are $s_0=0.1194$, $s_1=0.2122$, $s_2=-0.0671$ and $s_3=0.1031$. Galaxy and magnification biases are assumed to be constant in each redshift bin, with value determined at the mean redshift. 
We consider two configurations, with 5 and 10 equally spaced redshift bins in the redshift range $0.1 < z < 2$ with $\Delta z_5 =0.38$ for 5 bins and $\Delta z_{10} =0.19$ for 10 bins. The photometric redshift uncertainty is modelled with Gaussian bins with standard deviation $\Delta z_{i}/2$, $i=5,10$, for the observed power spectra in the errors $\sigma^{2}_{C_{\ell}}$ whereas we use top-hat bins with half-width $\Delta z_{i}/2$ for the theoretical power spectra in the derivatives with respect to the parameters in the Fisher matrix.

For the SKA spectroscopic survey we assume specifications as in \cite{Camera:2014bwa, Santos:2015hra} consistent with the 5 $\mu{\rm Jy}$ sensitivity:
\begin{align}
\frac{{\rm d}N}{{\rm d}z{d}\Omega}&=\left(\frac{180}{\pi^2}\right)10^{c_1}z^{c_2}  {\rm exp} \left(-c_3\right)\\
\text{for}\quad & 0.1 < z < 1.5 \\
f_{\rm sky}&=0.73 \\
b(z)&=b_0 {\rm exp} \left(b_1 z\right) \,, \label{b0SKA}
\end{align}
where $c_1=6.7767$, $c_2=2.1757$, $c_3=6.6874$, the fiducial value for the galaxy bias at $z=0$ is $b_0=0.5887$ and $b_1=0.8130$.

We compute the magnification bias according from the fitting formula for the cumulative number density of galaxies for SKA in Eq.~\eqref{nflux} as given in \cite{Camera:2014bwa}.
From Eq.~\eqref{magnbias} we find
\begin{equation}
s(z)=  s_0 +s_1z+s_2z^2+s_3z^3\,,
\end{equation}
with coefficients $s_0=-0.106875$, $s_1=1.35999$, $s_2=-0.620008$ and $s_3=0.188594$. Note that, physically, the magnification bias assumes positive values at all redshifts\footnote{We notify the reader that the fitting formula for the magnification bias for SKA appearing in Refs.~\cite{Raccanelli:2016avd, Raccanelli2015, Montanari:2015rga, DiDio:2016ykq} is incorrect. This is due an error in fitting formula for the cumulative number density of galaxies in Ref.~\cite{Camera:2014bwa} which was corrected afterwards.}. Galaxy and magnification biases are assumed to be constant in each redshift bin, with value determined at the mean redshift.

For SKA, we consider five configurations, with 5, 10, 20, 30 and 40 equally spaced redshift bins in the redshift range $0.1 < z < 1.5$. Given the spectroscopic redshift determination, we use top-hat redshift bins with half-width $\Delta z_{i}/2$, $i=5,10, 20, 30, 40$, both for the observed and for the theoretical power spectra in the Fisher matrix. 
For SKA we also consider the case where the cross-correlations between far redshift bins including lensing are neglected. To be specific we discard all cross-cross-correlations with separation $\Delta z > \Delta z_{5}=0.28$. The reason is twofold: firstly in this way our results mimic to some extend what one finds with a $P(k)$ analysis by neglecting long range correlations, even if we are using the angular power spectra. 
The second reason is that in this configuration we can completely trust the robustness of the method we use to calculate the bias in the parameter estimation and our results: as we remark above, the Fisher matrix method is based on the Taylor expansion of the likelihood and thus it is implicitly based on the assumption that the difference between the two models, here angular power spectra with or without lensing, is small. This is not the case for far cross-bin correlations with a spectroscopic survey, where the amplitude of the Newtonian spectra is completely negligible and the signal is completely dominated by the lensing convergence contribution. But it is true for cross-correlations between adjacent bins, at all scales.
To recover the 3-dimensional information with a spectroscopic survey one needs order of $10^2$ bins. Here we limit our analysis to a maximum of 40 redshift bins because of computational convenience.

%-------------------------------------------------------------------------------------------------------------------------------------------------------------------------------------------------------------------------------------------------------
%%%%%%%%%%%%%%%%%%%%%%%%%%%%%%%%%%%%%%%%%%%%%%%%%%%%%%%%%%%%%%%%%%%%%%%
%%%%%%%%%%%%%%%%%%%%%%%%%%%%%%%%%%%%%%%%%%%%%%%%%%%%%%%%%%%%%%%%%%%%%%%
\section{Results}
\label{sec:results}
%%%%%%%%%%%%%%%%%%%%%%%%%%%%%%%%%%%%%%%%%%%%%%%%%%%%%%%%%%%%%%%%%%%%%%%
%%%%%%%%%%%%%%%%%%%%%%%%%%%%%%%%%%%%%%%%%%%%%%%%%%%%%%%%%%%%%%%%%%%%%%%
The correlation between the cosmological parameters and the lensing parameter $\epsilon_L$ is one of the most important quantity to understand the behaviour of the effects of lensing that we want to study. It is defined by
\begin{equation}
\rho_{\epsilon_L}  \left( \theta_{\alpha} \right)= \frac{\left[\left(F^{\Phi\Phi}\right)^{-1}\right]_{\alpha\epsilon_L}}{\sqrt{\left[\left(F^{\Phi\Phi}\right)^{-1}\right]_{\alpha\alpha}\left[\left(F^{\Phi\Phi}\right)^{-1}\right]_{\epsilon_L\epsilon_L}}}\,,
\end{equation}
where the index $\alpha$ refers to the cosmological parameters $\{\theta_\alpha\}$ and we have included lensing in the Fisher matrix, both in the covariance and in the angular power spectra we use for the derivatives. The more this quantity is close to $1$ the more the correspondent parameter is correlated with lensing and therefore we would expect the impact of lensing to be more important for the constraints and for the shift of the best-value of that parameter. 

\begin{figure}[h!]
  \begin{subfigure}[b]{0.32\textwidth}
    \includegraphics[width=\textwidth]{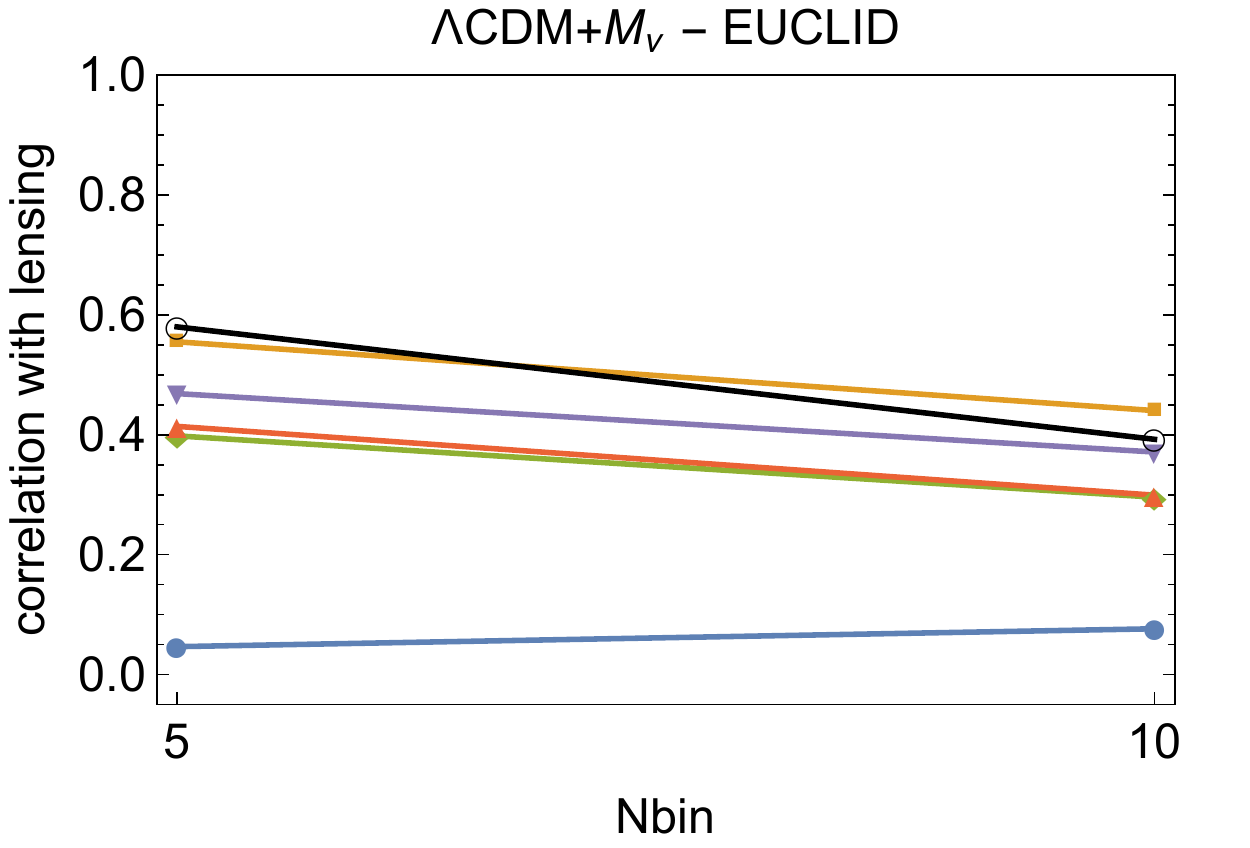}
   \end{subfigure}
 \hspace{-0.5cm}
    \begin{subfigure}[b]{0.32\textwidth}
    \includegraphics[width=\textwidth]{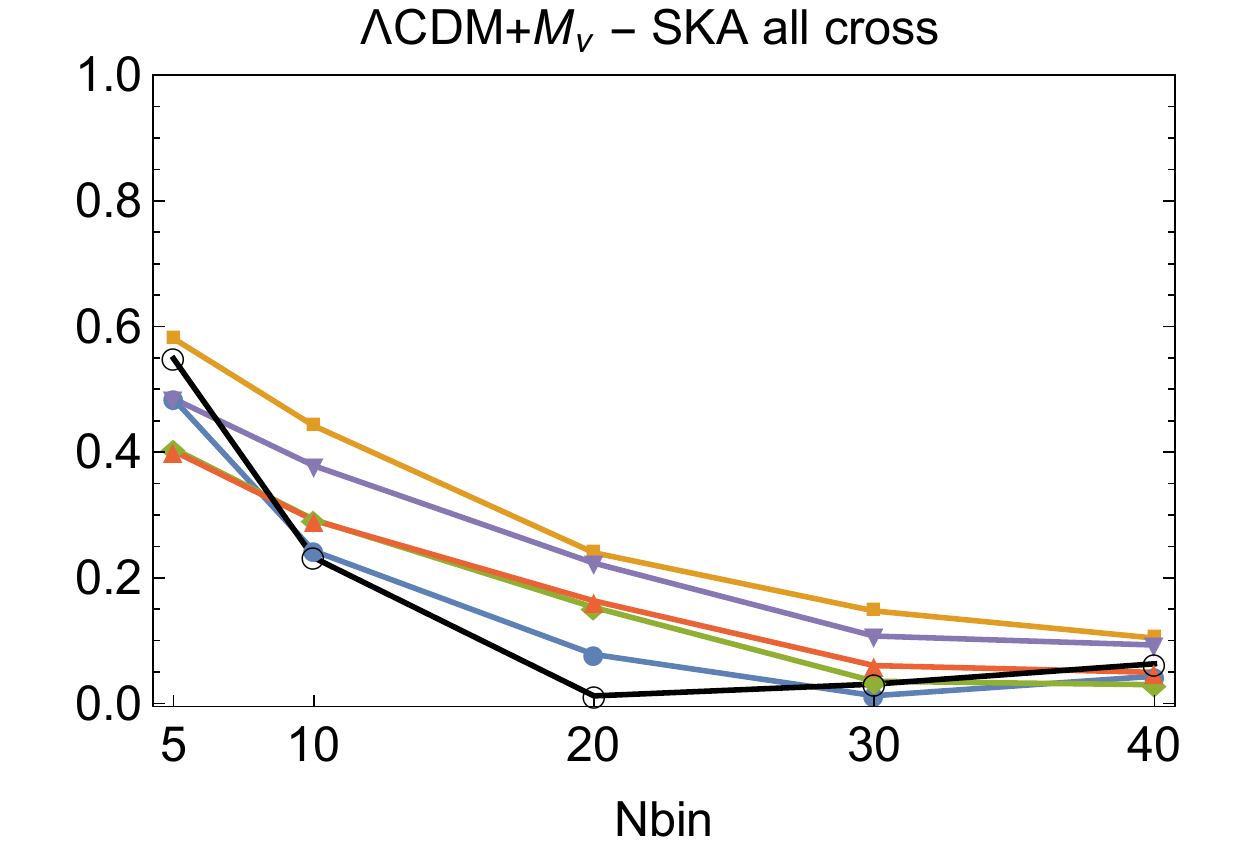}
   \end{subfigure}
 \hspace{-0.5cm}
    \begin{subfigure}[b]{0.32\textwidth}
    \includegraphics[width=\textwidth]{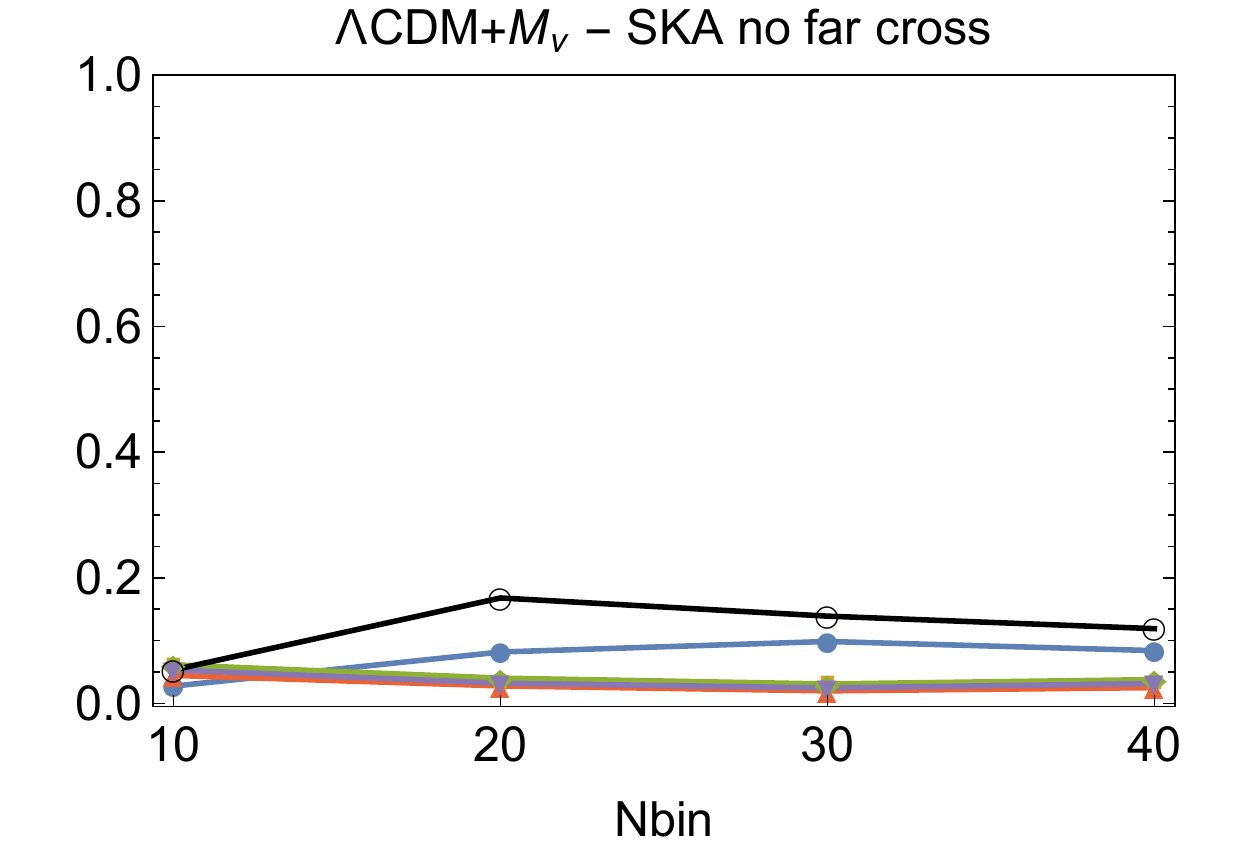}
   \end{subfigure}
 \hspace{-0.5cm}
   \begin{subfigure}[t]{0.065\textwidth}
   {\vspace{-3cm}
    \includegraphics[width=\textwidth]{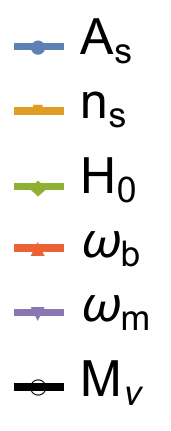}}
  \end{subfigure}
  \caption{The correlation of the lensing parameter $\epsilon_L$ with the parameters of the $\Lambda$CDM model with massive neutrinos. The correlation is given for different binning configurations for Euclid (left panel) and for SKA considering all the redshift bin cross-correlations (middle panel) and neglecting the cross-correlations between far redshift bins (right panel).}
   \label{fig:lenscorrLCDMn}
\end{figure}

\begin{figure}[h!]
  \begin{subfigure}[b]{0.32\textwidth}
    \includegraphics[width=\textwidth]{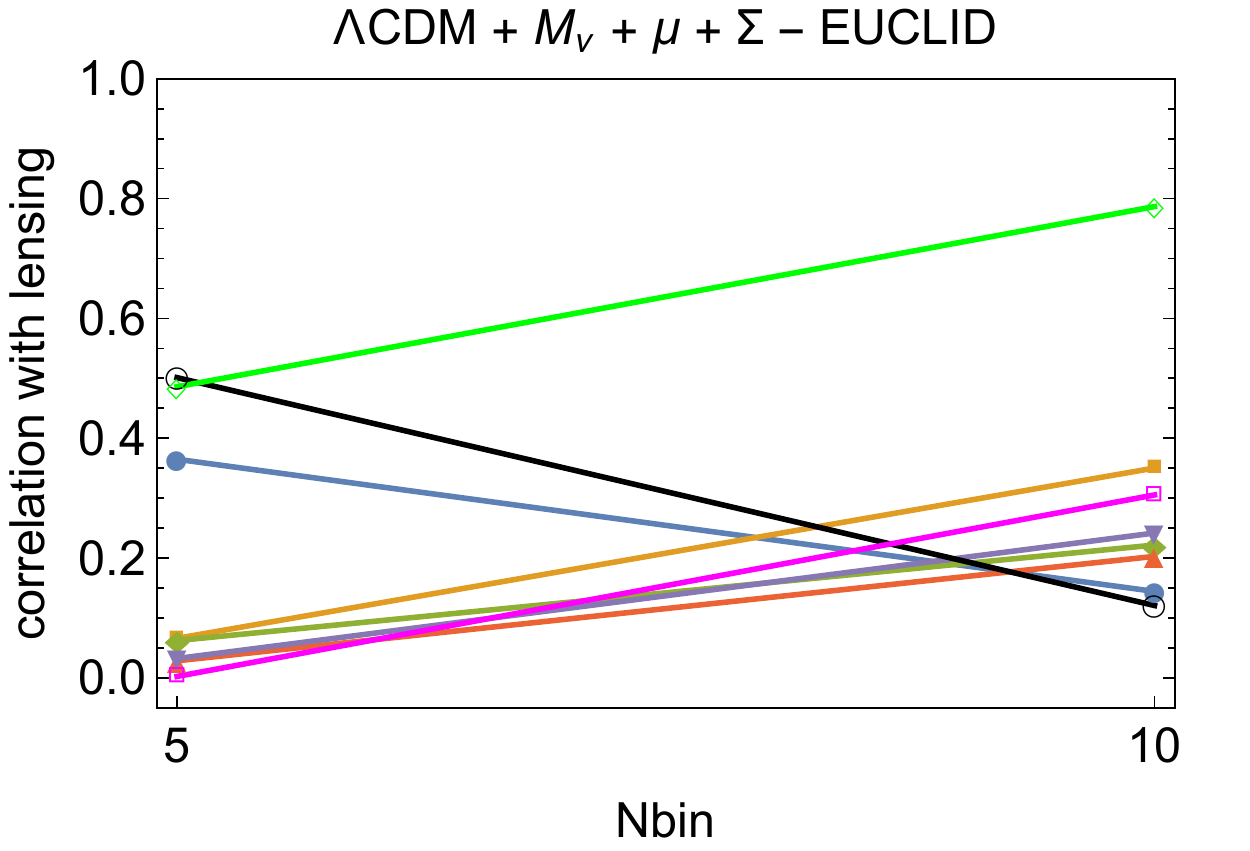}
   \end{subfigure}
 \hspace{-0.5cm}
    \begin{subfigure}[b]{0.32\textwidth}
    \includegraphics[width=\textwidth]{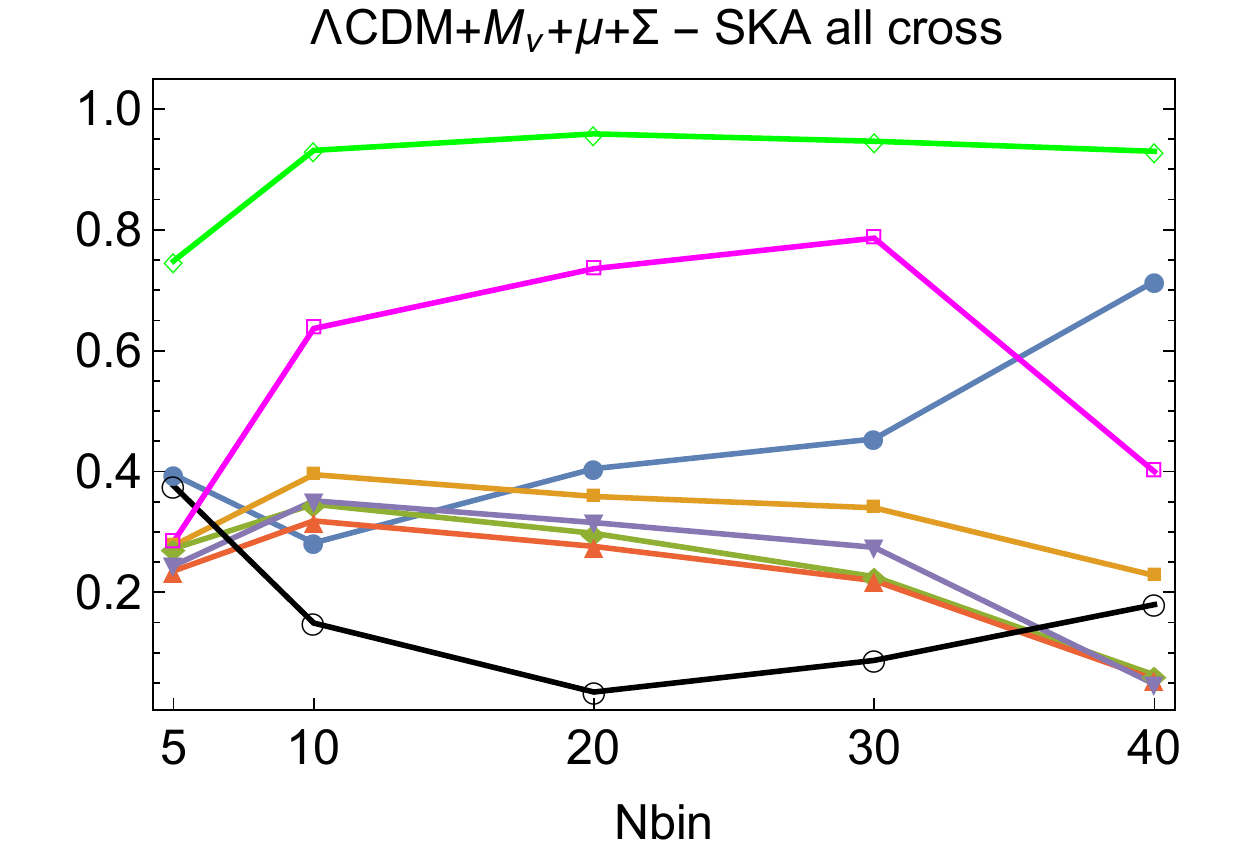}
   \end{subfigure}
 \hspace{-0.5cm}
    \begin{subfigure}[b]{0.32\textwidth}
    \includegraphics[width=\textwidth]{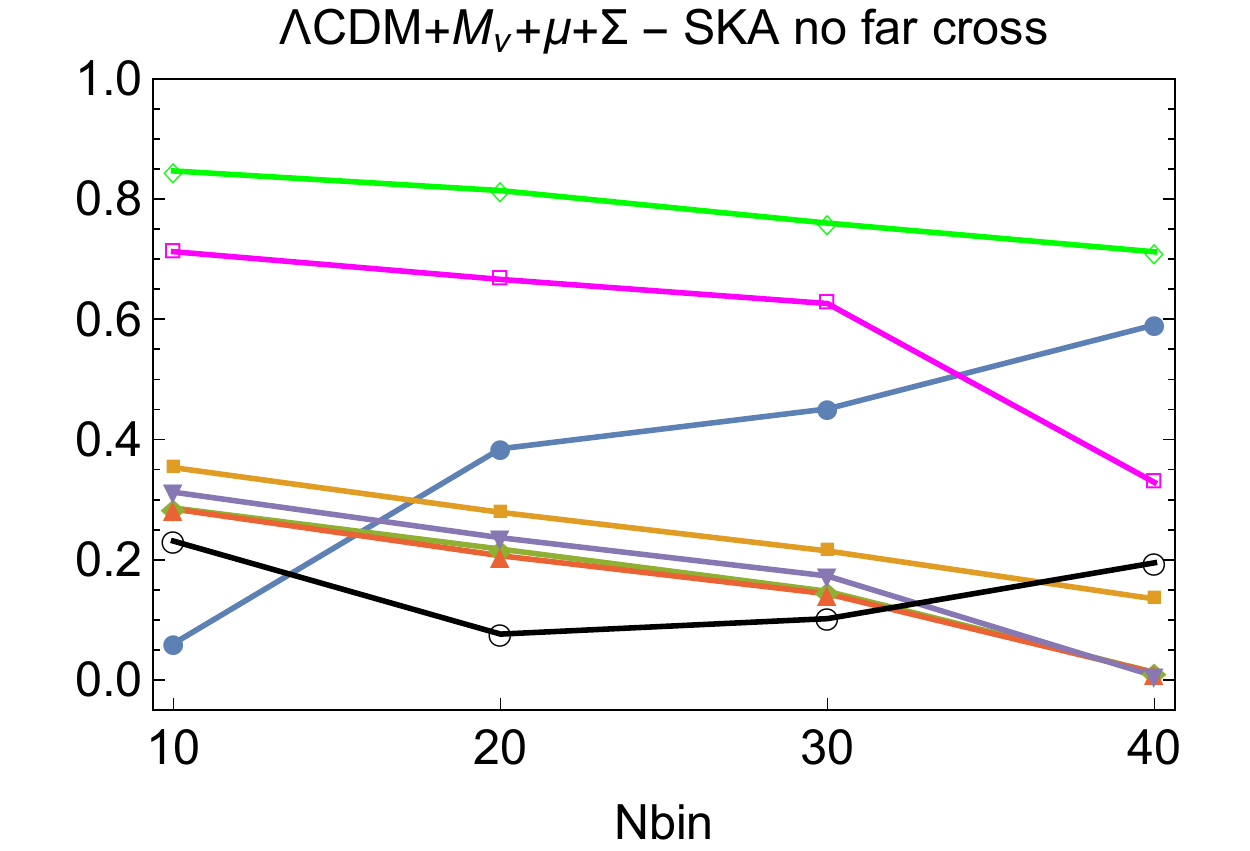}
   \end{subfigure}
 \hspace{-0.05cm}
   \begin{subfigure}[t]{0.065\textwidth}
    \includegraphics[width=\textwidth]{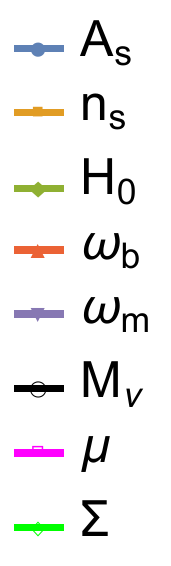}
  \end{subfigure}
  \caption{The correlation of the lensing parameter $\epsilon_L$ with the parameters of the $\Lambda$CDM model including massive neutrinos and modified gravity. The correlation is given for different binning configurations for Euclid (left panel) and for SKA considering all the redshift bin cross-correlations (middle panel) and neglecting the cross-correlations between far redshift bins (right panel).}
   \label{fig:lenscorrLCDMnMG}
\end{figure}

In Fig.~\ref{fig:lenscorrLCDMn} and in Fig.~\ref{fig:lenscorrLCDMnMG} we plot this correlation for the $\Lambda$CDM model including massive neutrinos and including also modified gravity, respectively. By looking first at Fig.~\ref{fig:lenscorrLCDMn} we see that all the $\Lambda$CDM parameters, including the sum of neutrino masses, are all not much sensitive to lensing.
From the results for SKA when we include all the possible cross-correlations we also see that the correlation with $\epsilon_L$ decreases as the number of redshift bins increases, as one would expect since the lensing signal becomes relatively weaker. Indeed, by increasing the number of bins, we consider more radial modes which are dominated by the standard Newtonian contributions, being the lensing a pure transversal effect. In addition, in the case where we neglect far-bin cross-correlations the correlation with $\epsilon_L$ is in general smaller for all the parameters with values which do not exceed $0.2$ and are independent of the number of bins.
When we include modified gravity, the correlation of the $\Lambda$CDM parameters and the neutrino mass $M_\nu$ behaves in the same way as for the $\Lambda$CDM + $M_\nu$ case, with the only exception of the amplitude of scalar perturbations $A_s$, which becomes more correlated with lensing especially for SKA, as shown in Fig.~\ref{fig:lenscorrLCDMnMG}. The parameter $\Sigma$ which parametrizes the sum of the two Bardeen potentials in modified gravity (see Eq.~\eqref{sigma}) is the parameter most correlated with lensing, as one would expect since lensing is crucial to probe this quantity. This can be clearly seen from Fig.~\ref{fig:lenscorrLCDMnMG}, especially for SKA, where the correlation between $\Sigma$ and $\epsilon_L$ stays constant and close to 1 as the number of redshift bins increases (middle panel of Fig.~\ref{fig:lenscorrLCDMnMG}), also for the configuration where we drop the far-bin correlations (right panel of Fig.~\ref{fig:lenscorrLCDMnMG}). Note that as the number of redshift bins increases the correlations with $\epsilon_L$ for SKA including all the cross-correlations and those neglecting far-bin cross-correlations become almost the same, for all the three models we consider. This indicates that as one increases the number of bins the correlations with lensing is not controlled by the far-bin cross-spectra, despite the lensing signal dominates over the Newtonian terms. 
This is a first indication that for most of the cosmological parameters, lensing magnification is not carrying crucial information for a spectroscopic survey.

%%%%%%%%%%%%%%%%%%%%%%%%%%%%%%%%%%%%%%%%%%%%%%%%%%%%%%%%%%%%%%%%%%%%%%%
\subsection{The effect of lensing magnification: constraints on the cosmological parameters}
%%%%%%%%%%%%%%%%%%%%%%%%%%%%%%%%%%%%%%%%%%%%%%%%%%%%%%%%%%%%%%%%%%%%%%%
\label{sec:Rconstraints}
In this section we quantify how much information is present in the lensing contribution in galaxy clustering.
Our aim here is not to provide precise forecast constraints, rather to estimate the relative impact of lensing compared to the standard Newtonian analysis.
We therefore present our results in terms of the ratio of the marginalized errors calculated with two different Fisher analyses, the difference being the contributions we consider in the model for the galaxy power spectra: intrinsic clustering and redshift-space distortions only in one case and including also lensing convergence in the other case. If this ratio is close to $1$ then the constraining power of lensing is negligible. On the contrary, for values significantly less than $1$, the improvement on the errors coming from lensing is important.
The errors are marginalized over the entire set of the parameters for each model we consider: we have $5$ parameters for $\Lambda$CDM, $6$ parameters when we include massive neutrinos and $8$ parameters if we also include modified gravity. We address to section~\ref{sec:Mconstraints} for a detailed discussion of our approach. 
We also analyse the dependence of the constraining power of lensing on the number of redshift bins. The relative behaviour of the errors on a parameter - with or without lensing - is due to the dependence of the lensing magnification on that parameter and is also due to the dependence of the lensing signal on the survey and the binning configuration.

\begin{figure}[t]
   \begin{subfigure}[b]{0.32\textwidth}
    \includegraphics[width=\textwidth]{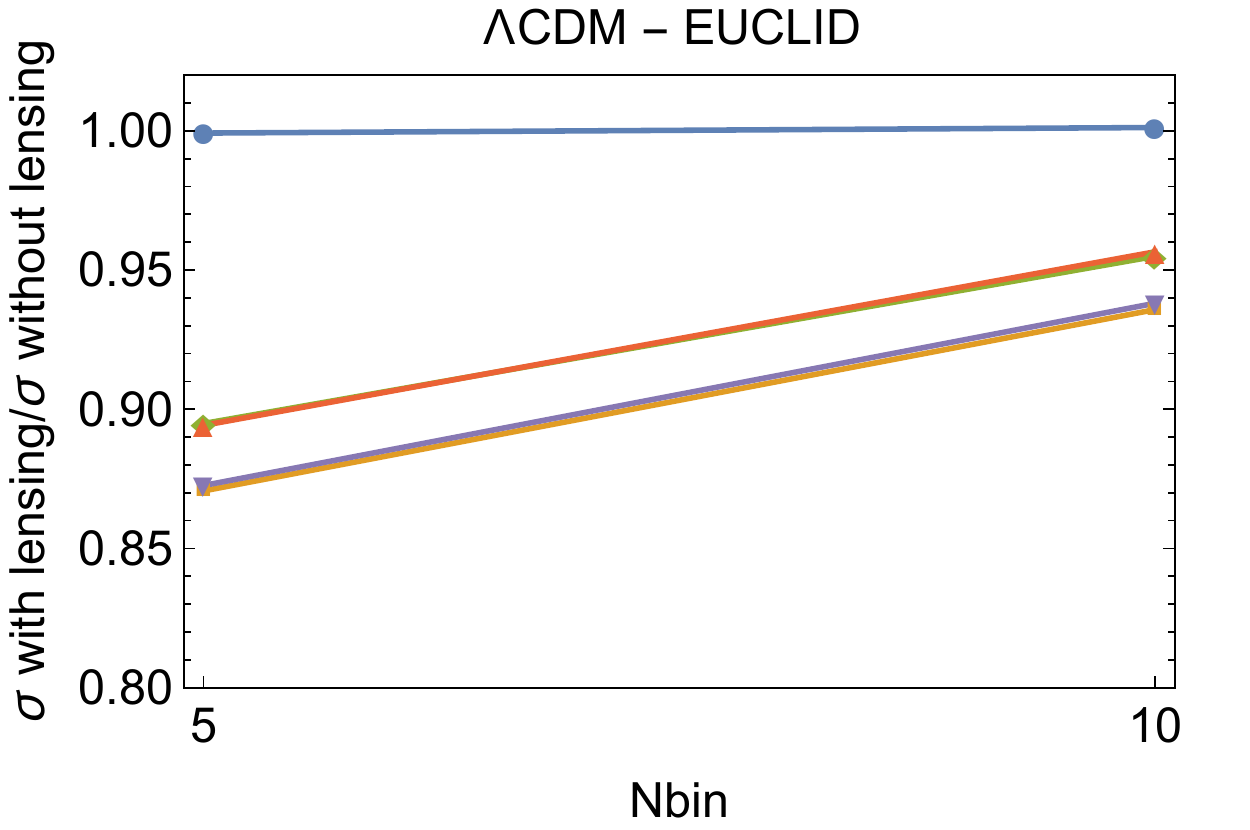}
    %\label{fig:sigmaLCDMEU}
  \end{subfigure}
 \hspace{-0.5cm}
  \begin{subfigure}[b]{0.32\textwidth}
    \includegraphics[width=\textwidth]{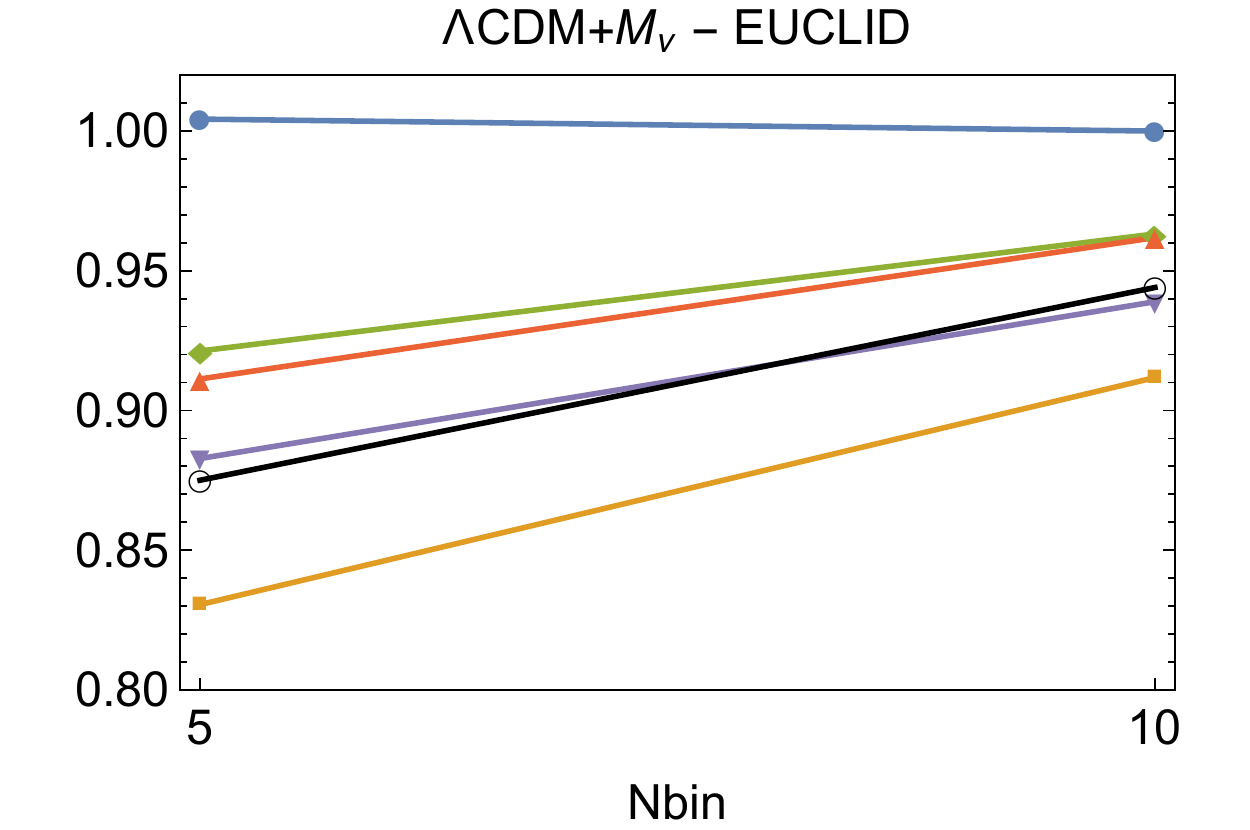}
    %\label{fig:sigmaLCDMnEU}
  \end{subfigure}
 \hspace{-0.5cm}
  \begin{subfigure}[b]{0.32\textwidth}
    \includegraphics[width=\textwidth]{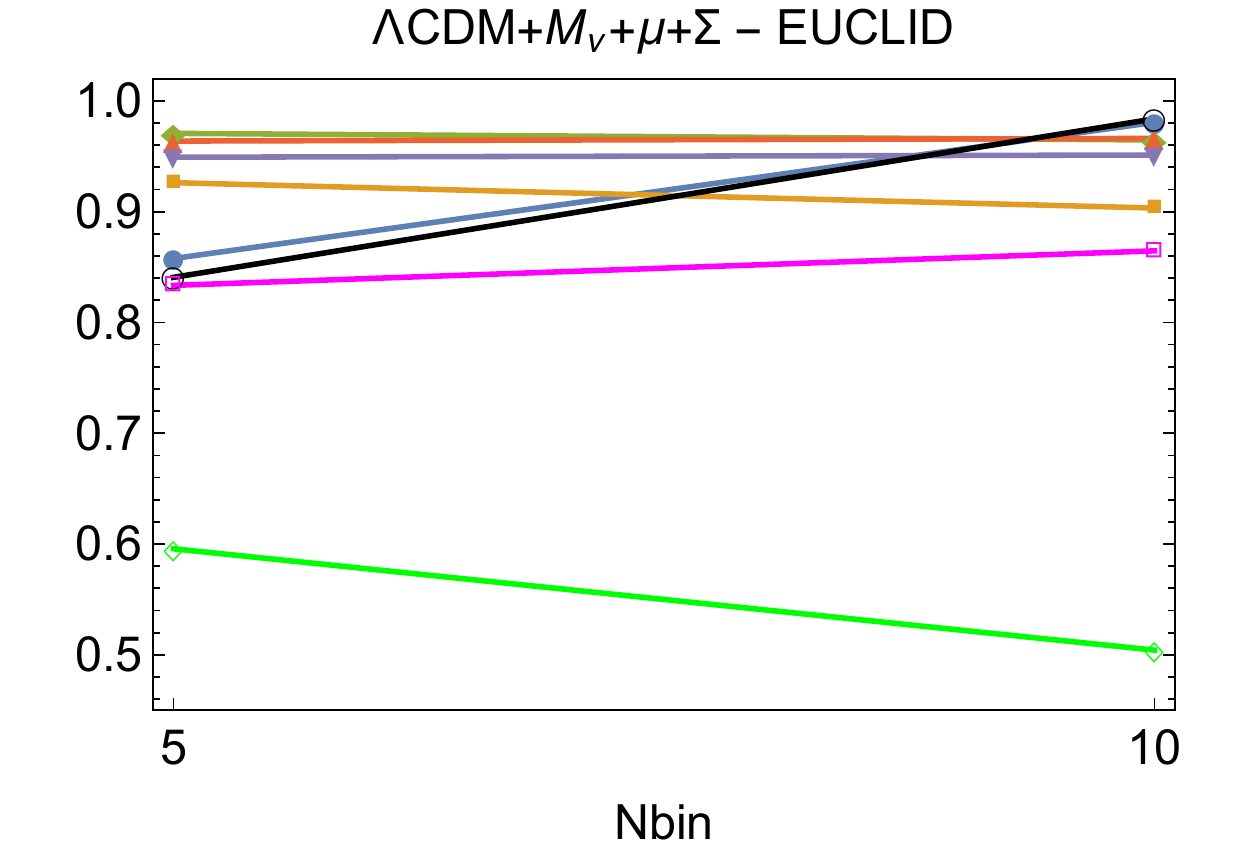}
  %  \label{fig:sigmaLCDMnMGEU}
  \end{subfigure}
  \hspace{-0.05cm}
  \begin{subfigure}[b]{0.065\textwidth}
  \includegraphics[width=\textwidth]{mblegendsigma.pdf}
  \end{subfigure}
   \caption{Results for Euclid: the ratio of the errors on the cosmological parameters calculated including lensing to the ones calculated neglecting it. In both analyses we include lensing contribution in the covariance matrix.}
   \label{fig:sigmaEU}
  \end{figure}

\begin{figure}[t]
   \begin{subfigure}[b]{0.32\textwidth}
    \includegraphics[width=\textwidth]{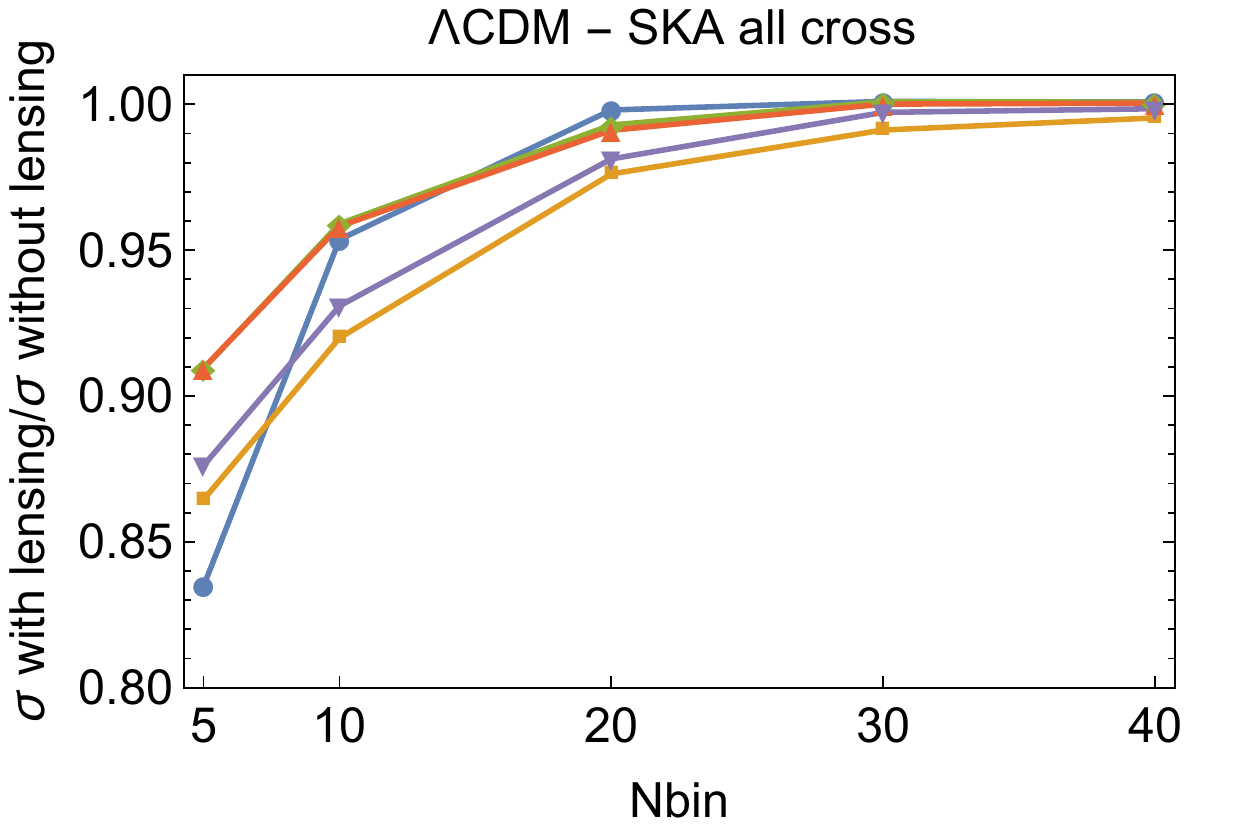}
    \end{subfigure}
  \hspace{-0.5cm}
  \begin{subfigure}[b]{0.32\textwidth}
  \includegraphics[width=\textwidth]{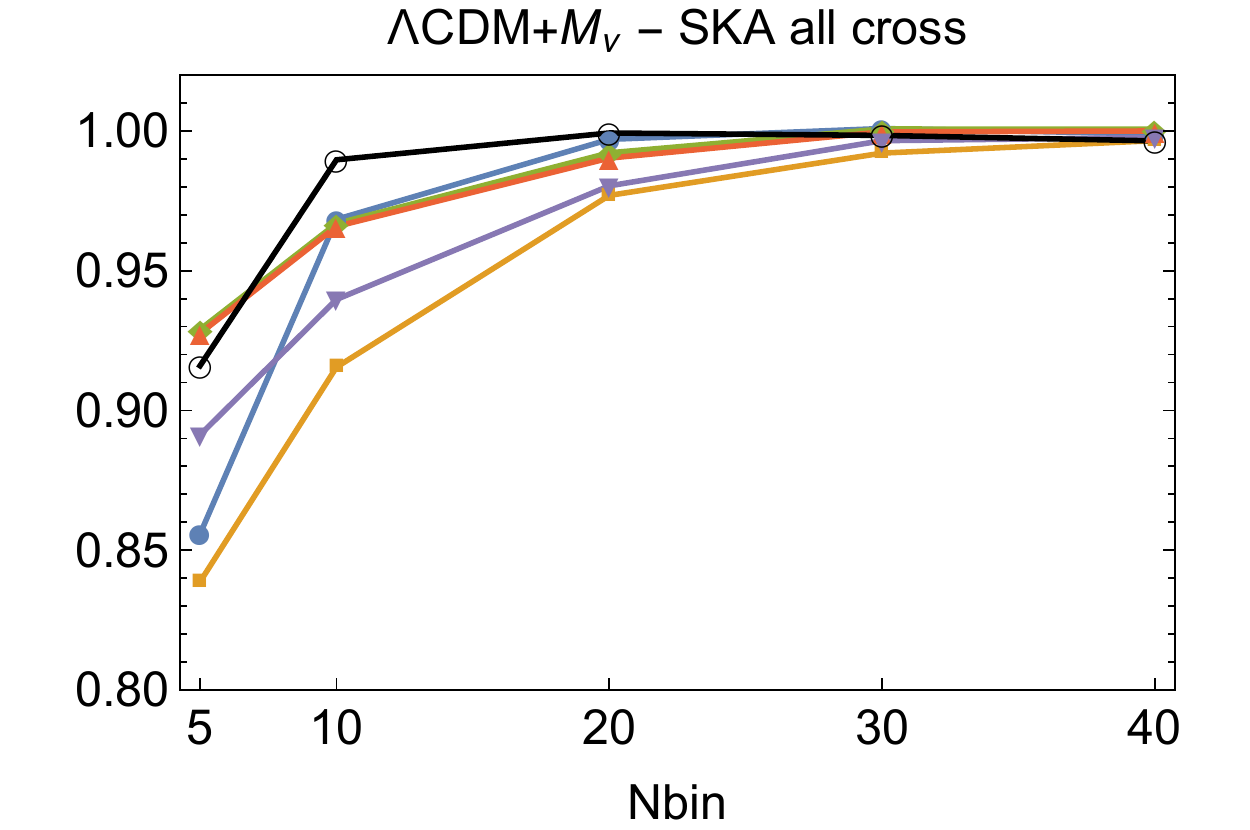}
  \end{subfigure}
   \hspace{-0.5cm}
  \begin{subfigure}[b]{0.32\textwidth}
    \includegraphics[width=\textwidth]{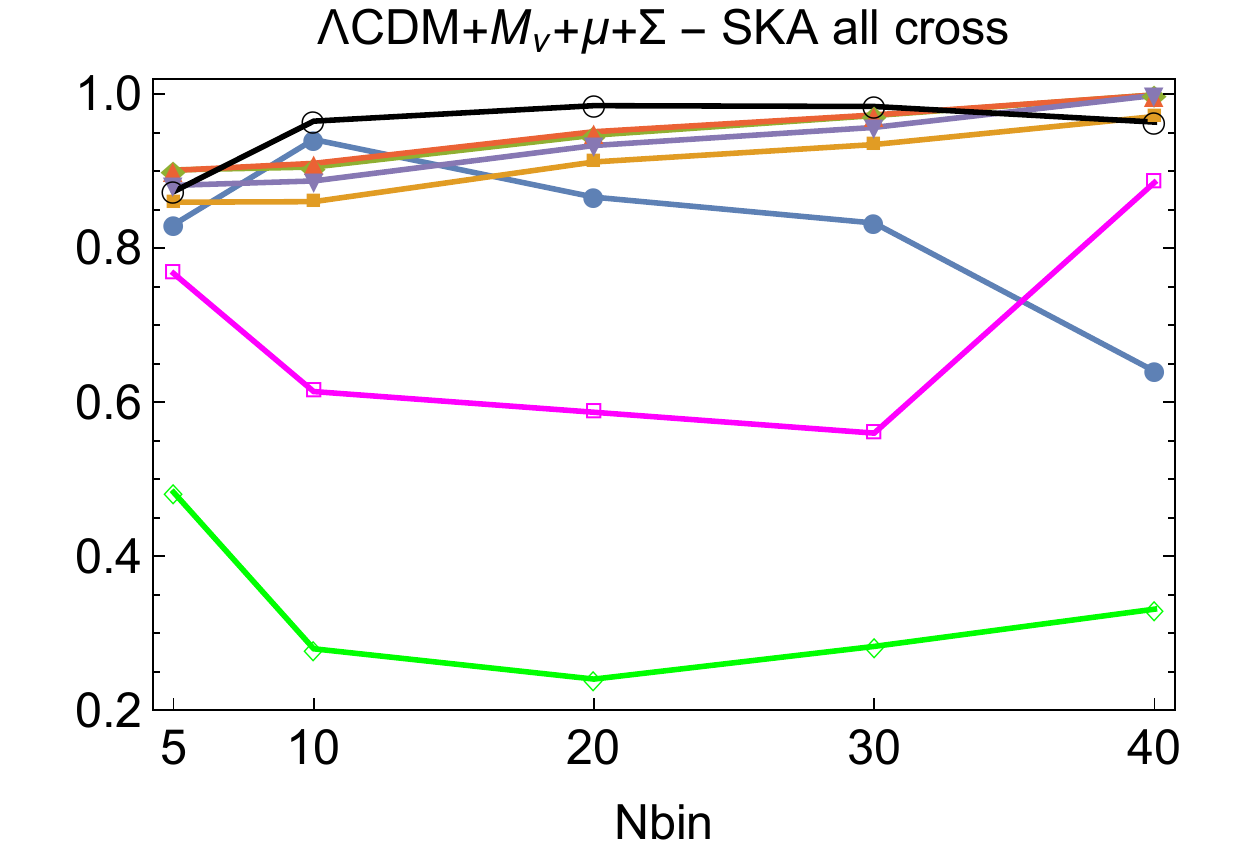}
   \end{subfigure}
   \hspace{-0.05cm}
    \begin{subfigure}[b]{0.065\textwidth}
    \includegraphics[width=\textwidth]{mblegendsigma.pdf}
  \end{subfigure}
  \caption{Results for SKA: the ratio of the errors on the cosmological parameters calculated including lensing to the ones calculated neglecting it. In both analyses we include lensing contribution in the covariance matrix.}
   \label{fig:sigmaSKA}
\end{figure}

Fig.~\ref{fig:sigmaEU} shows the results for Euclid and Fig.~\ref{fig:sigmaSKA} for SKA.
%%%%%%%%%%%%%%
%%%%%%%%%%%%%%
%%%%%%%%%%%%%%%%%%%%
%%%%%%%%%%%%%%%%%%%%
Again we look first to the $\Lambda$CDM model and its extension with massive neutrinos. From left and middle panels in Fig.~\ref{fig:sigmaEU} and in Fig.~\ref{fig:sigmaSKA} we see that the information contained in the lensing contribution does not improve substantially the constraints: the ratios of the errors for all the parameters rapidly approach 1 as the number of redhsift bins increases. The results are also shown in Table~\ref{tableLCDMn} in Appendix~\ref{sec:table}.
In the right panel of Fig.~\ref{fig:sigmaEU} and~\ref{fig:sigmaSKA} we show the results for modified gravity. We see that lensing improves the constraints on the modified gravity parameters, the strongest improvement being on $\Sigma$: for Euclid with $10$ bins the error with lensing is one half the error without lensing and for SKA with $40$ bins and considering all the cross-correlations the ratio of the errors is $0.33$, see Table~\ref{tableLCDMnMG} in Appendix~\ref{sec:table}.
In addition, for SKA the right panel of Fig.~\ref{fig:sigmaSKA} tells us that adding lensing magnification in modified gravity improves the constraints on the amplitude of primordial fluctuations $A_s$, because the lensing potential breaks degeneracy between these parameters.

It is worth discussing in more detail the behaviour of the relative error of the modified gravity parameters $\mu$ and $\Sigma$ for the case of the spectroscopic survey SKA. By inspecting the right panel of Fig.~\ref{fig:sigmaSKA} we note that the improvement on the error of $\Sigma$ is significant for the $30$ and $40$ bins configuration, the ratio of the errors with and without lensing being lower than $0.4$. On the contrary for the parameter $\mu$ the ratio of the errors pass from $0.5598$ at $30$ bins to $0.8853$ at $40$ bins - approaching the value of the other $\Lambda$CDM parameters - indicating that the constraining power of lensing on $\mu$ decreases as the number of redshift bins increases.
Indeed by increasing the number of redshift bins, we start considering shorter radial modes. With a better radial resolution we resolve the redshift space distortion contributions. This leads to a measurement of the growth rate and it provides constraints on the galaxy clustering parametrized by $\mu$. Therefore the constraints on $\mu$ are driven by the standard newtonian contributions in the limit of a spectroscopic survey.

\begin{figure}[h!]
\begin{subfigure}[h!]{0.46\textwidth}
    \includegraphics[width=\textwidth]{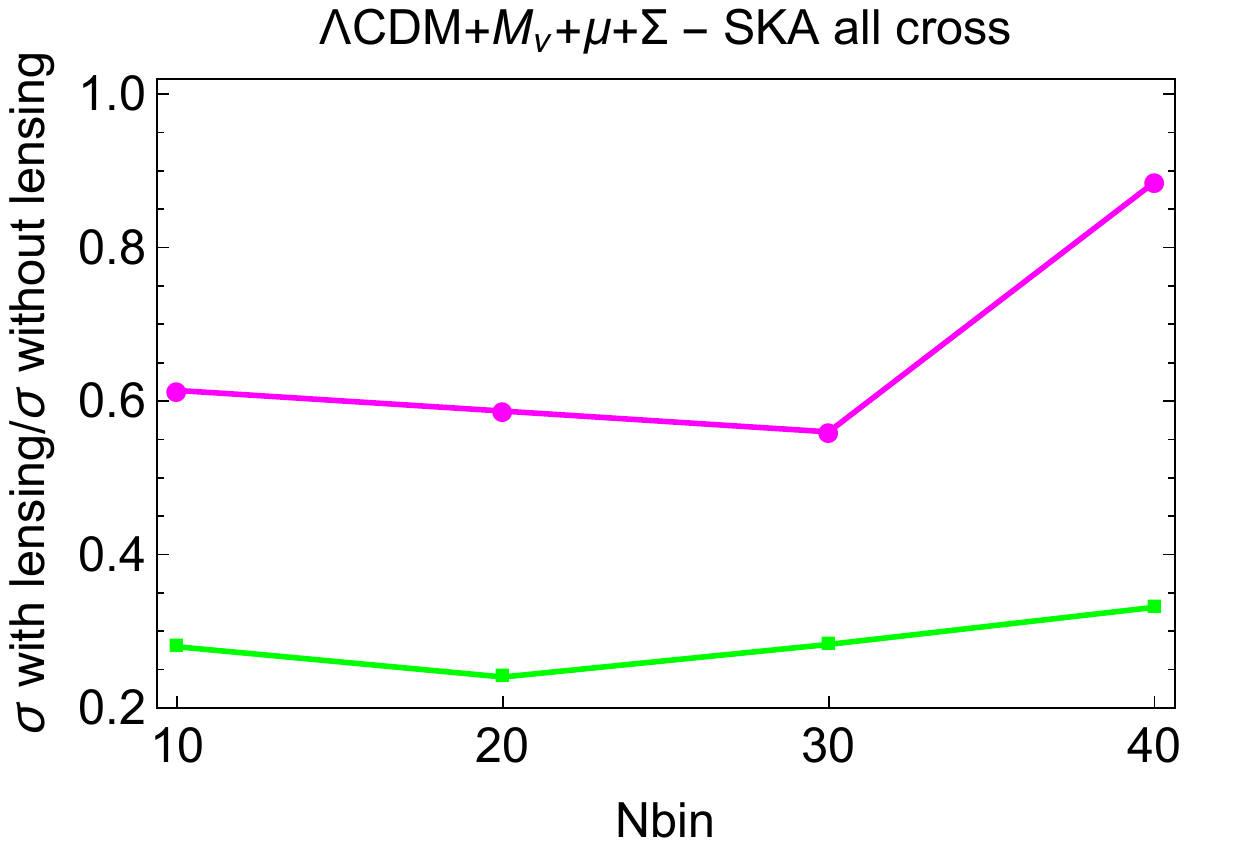}
  \end{subfigure}
  \begin{subfigure}[h!]{0.54\textwidth}
    \includegraphics[width=\textwidth]{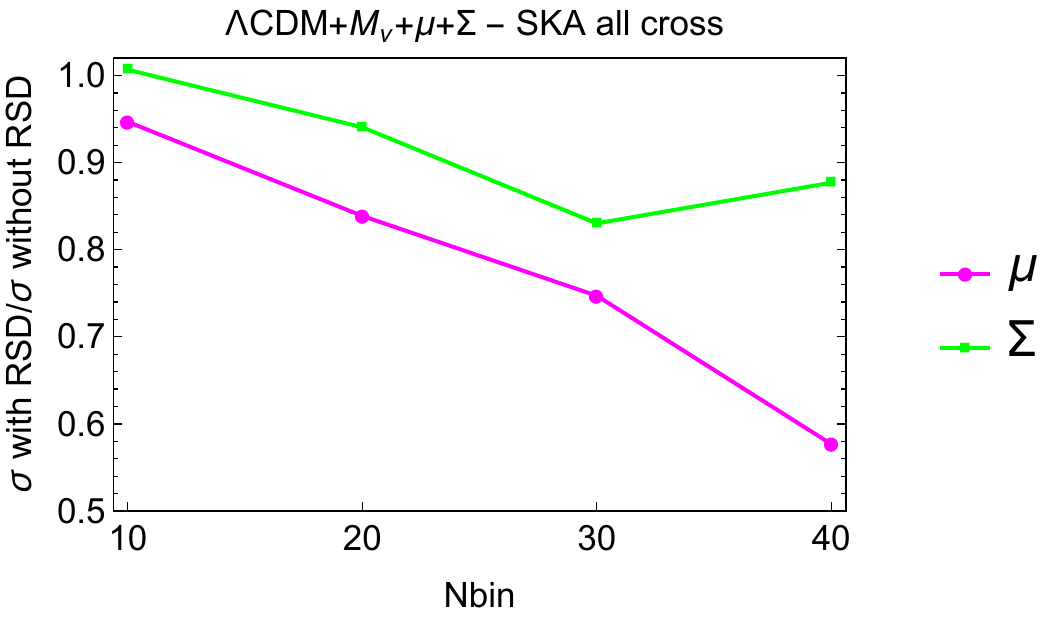}
   \end{subfigure} 
    \caption{
    Left panel: the ratio of the errors on the modified gravity parameters calculated including lensing to the ones calculated neglecting it. 
    Right panel: the ratio of the errors on the modified gravity parameters calculated including density and redshift-space distortion to the ones calculated with density alone.
    The errors are calculated by marginalizing over the other parameters of the $\Lambda {\rm CDM} +M_\nu + \mu + \Sigma$ model. Here we consider all the cross-correlation between the redshift bins for SKA.}
    \label{fig:sigmaRSD}
  \end{figure}

To corroborate this statement we isolate the effect of redshift-space distortions by performing an additional Fisher analysis to examine their constraining power on the modified gravity parameters. In Fig.~\ref{fig:sigmaRSD} we show on the right panel the ratio of the errors calculated including density and redshift-space distortion to the ones calculated with density alone and we report on the left panel the ratio of the errors given by including to neglecting lensing. It is clear the different behaviour of the two modified gravity parameters. From the right panel we see that redshift-space distortions improve the error on $\mu$ whereas they do not add useful information for $\Sigma$ starting from 30 bins, the ratio of the errors on $\Sigma$ becoming closer to 1 for 40 bins. On the contrary, from the left panel we see the opposite behaviour: starting from 30 bins adding lensing stops improving the error on $\mu$ whereas the improvement of the errors on $\Sigma$ stay roughly constant by increasing the number of bins.
Our results also validate the choice of the parametrization $\{ \mu, \Sigma \}$ to test deviation from GR on large scales. Indeed with this choice we decouple the effect induced by structure formation on $\mu$ and the lensing effect on $\Sigma$.

%%%%%%%%%%%%%%%%%%%%%%%%%%%%%%%%%%%%%%%%%%%%%%%%%%%%%%%%%%%%%%%%%%%%%%%
\subsection{The effect of lensing magnification: bias on the cosmological parameters estimation}
%%%%%%%%%%%%%%%%%%%%%%%%%%%%%%%%%%%%%%%%%%%%%%%%%%%%%%%%%%%%%%%%%%%%%%%
\label{sec:Rshifts}

\begin{figure}[h!]
  \begin{subfigure}[h!]{0.345\textwidth}
    \includegraphics[width=\textwidth]{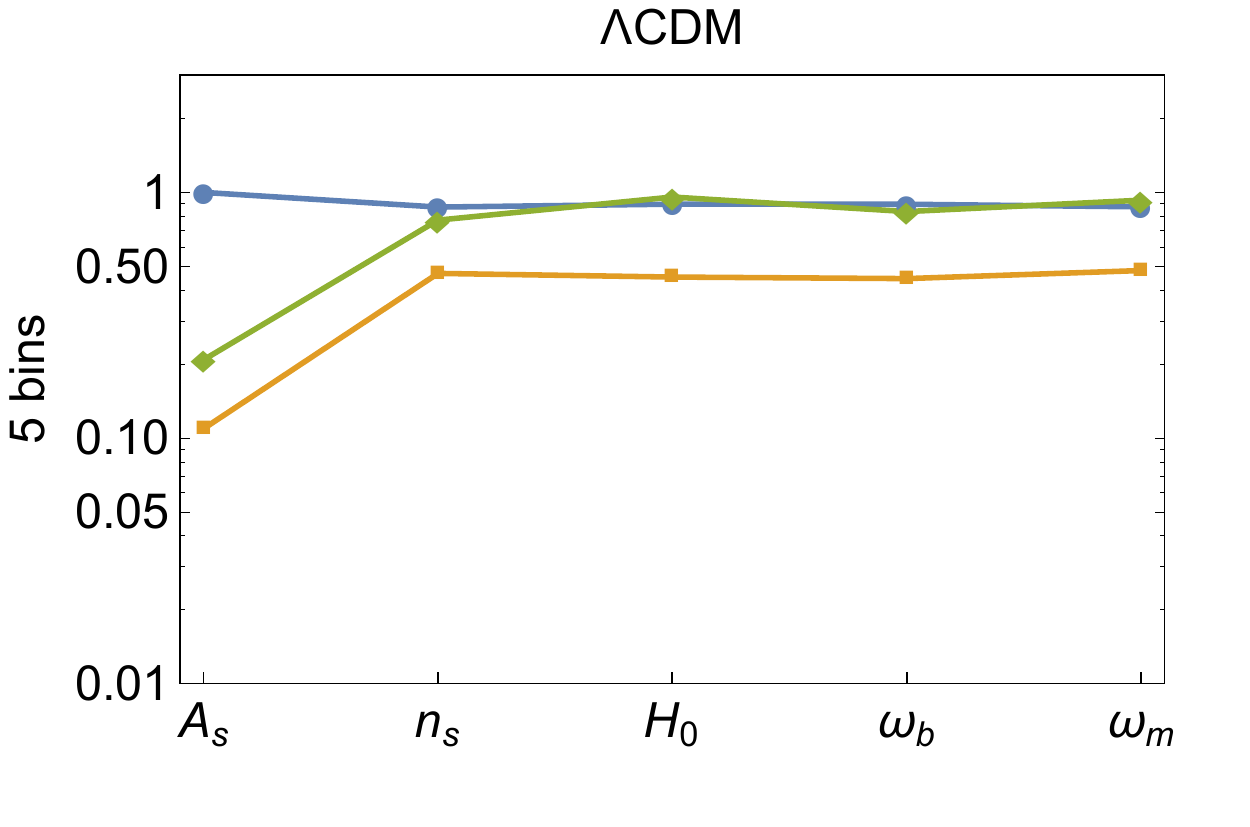}
    \label{fig:corrLCDMEU5}
  \end{subfigure}
 \hspace{-2.5cm}
  \begin{subfigure}[h!]{0.345\textwidth}
    \includegraphics[width=\textwidth]{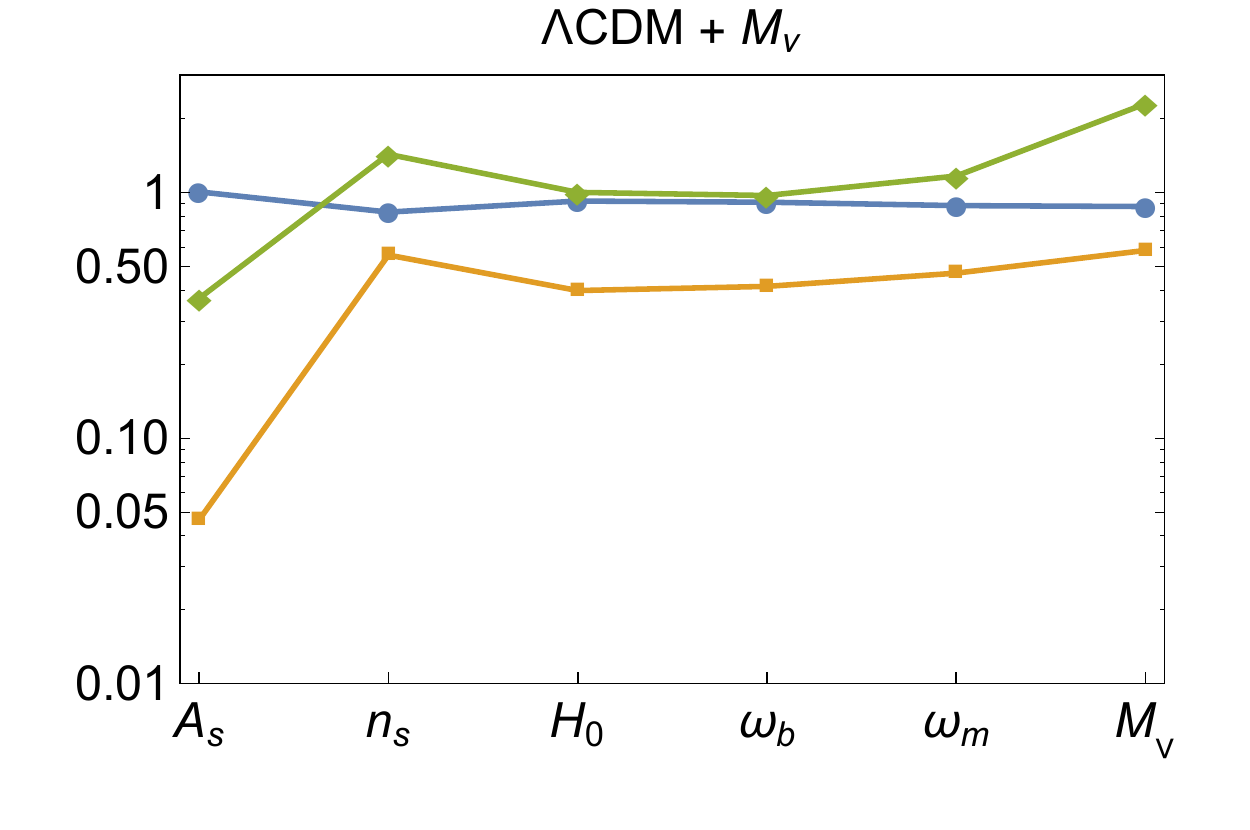}
    \label{fig:corrLCDMnEU5}
  \end{subfigure}
 \hspace{-2.5cm}
  \begin{subfigure}[h!]{0.345\textwidth}
    \includegraphics[width=\textwidth]{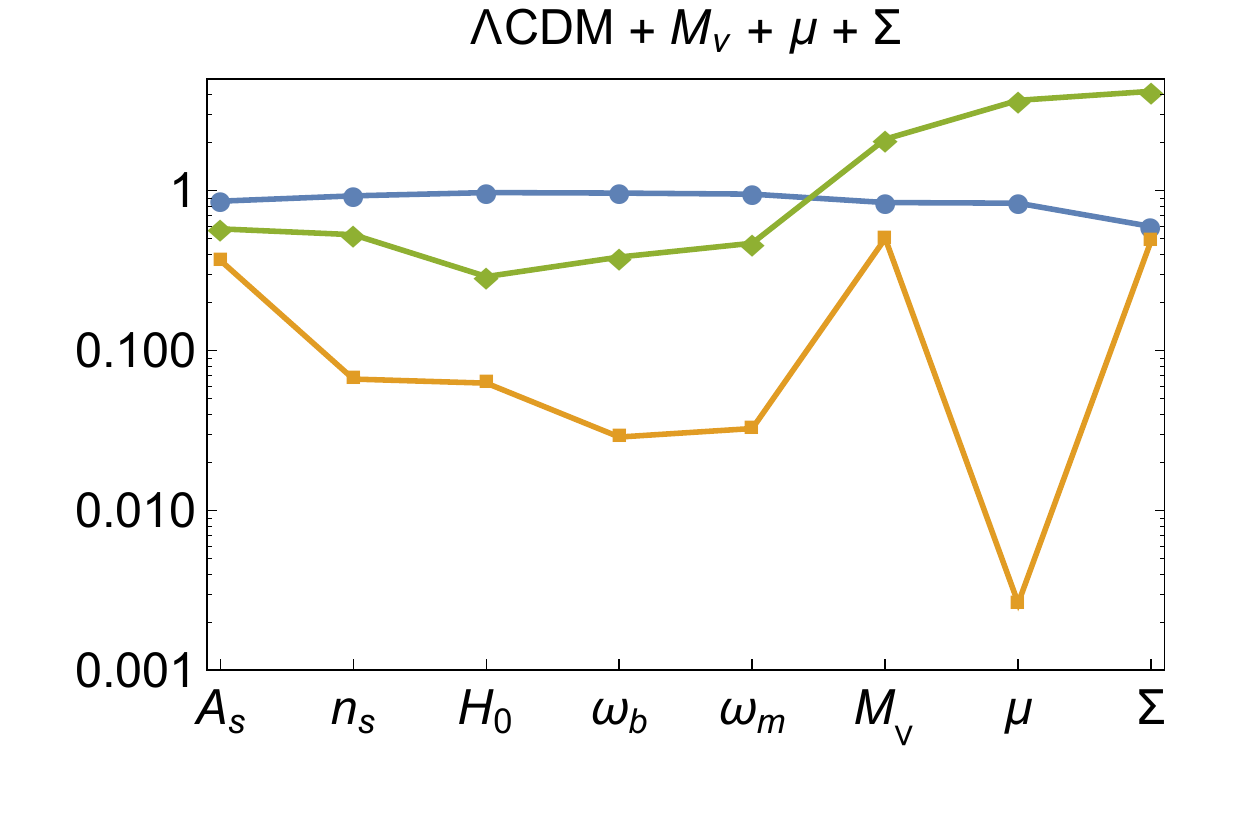}
    \label{fig:corrLCDMnMGEU5}
  \end{subfigure}
  %%%%%%%%%%%%%%%%%%%%%%%%%%%%%
  \begin{subfigure}[h!]{0.345\textwidth}
    \includegraphics[width=\textwidth]{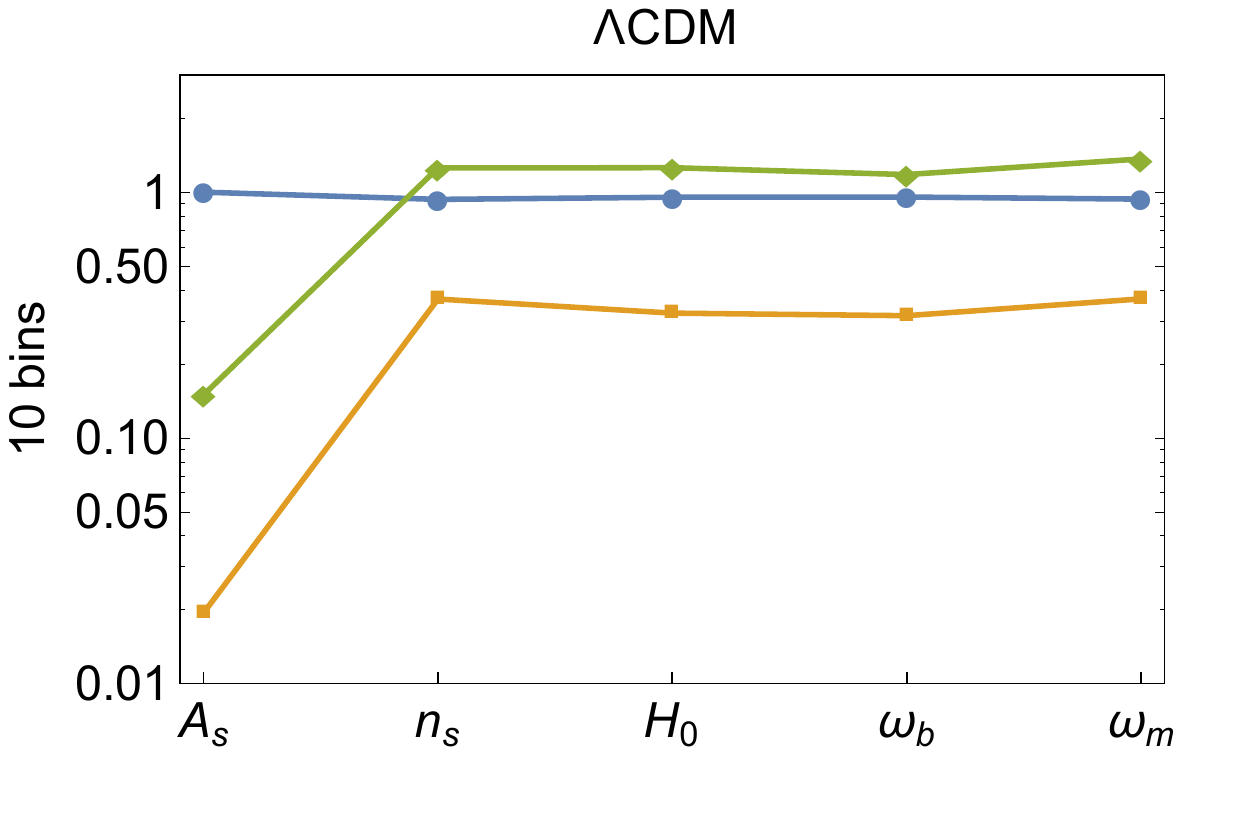}
    \label{fig:corrLCDMEU5}
  \end{subfigure}
 \hspace{-0.5cm}
  \begin{subfigure}[h!]{0.345\textwidth}
    \includegraphics[width=\textwidth]{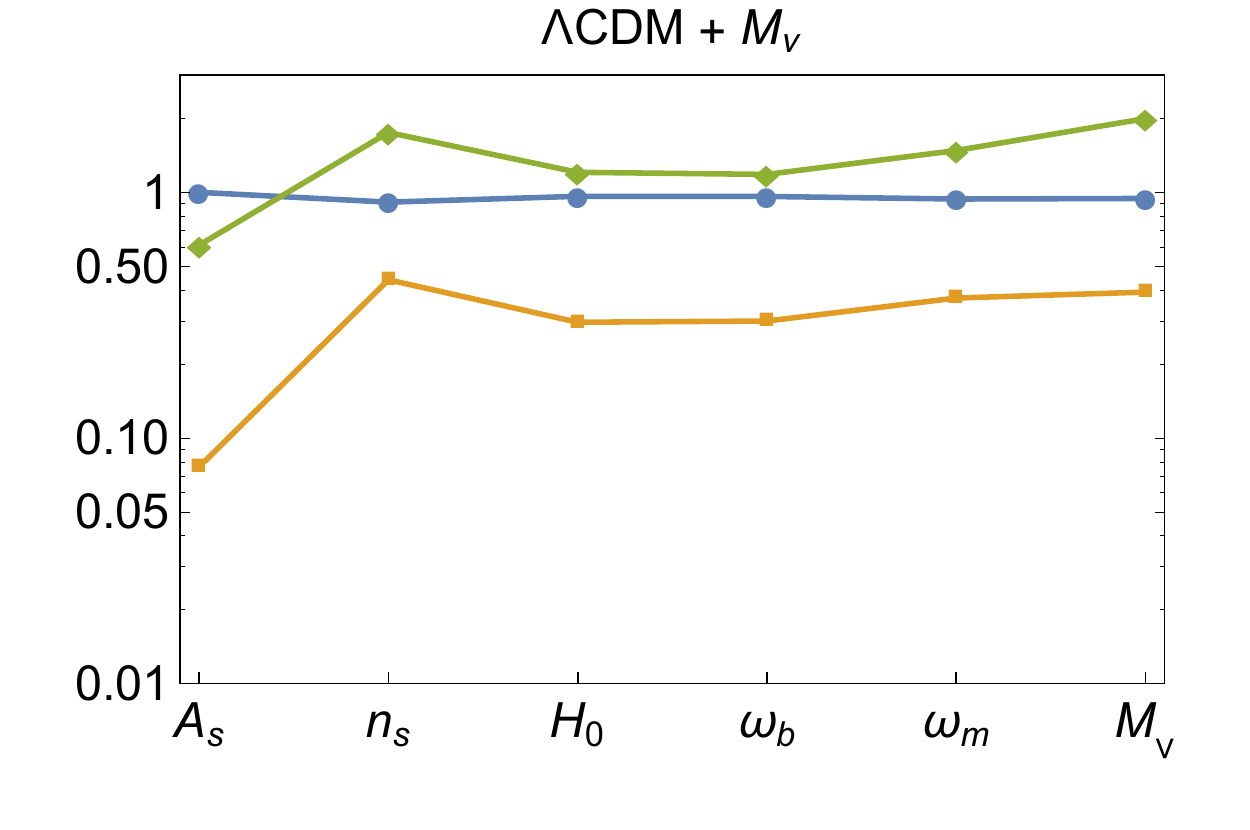}
    \label{fig:corrLCDMnEU5}
  \end{subfigure}
 \hspace{-0.5cm}
  \begin{subfigure}[h!]{0.345\textwidth}
    \includegraphics[width=\textwidth]{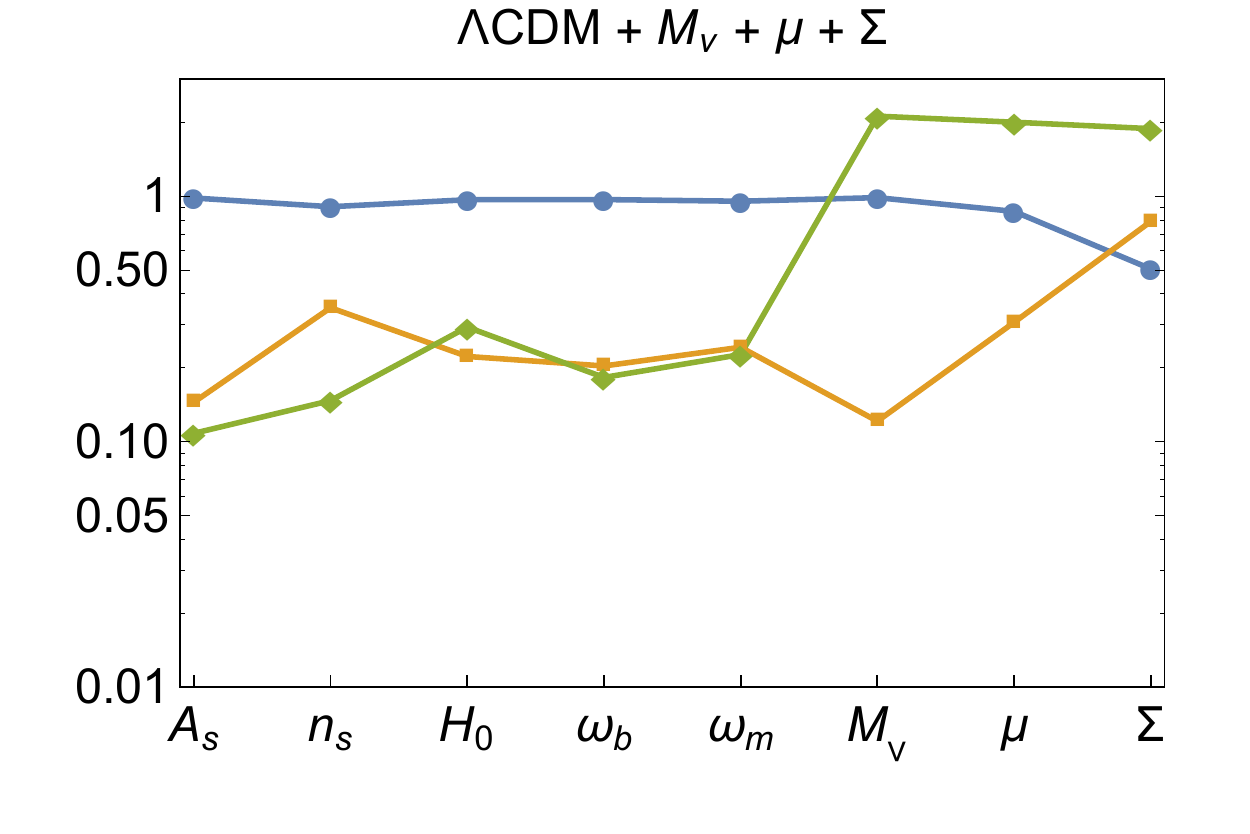}
    \label{fig:corrLCDMnMGEU10}
  \end{subfigure}
  \vspace{-0.8cm}
  \begin{center}
   \begin{subfigure}[h!]{0.3\textwidth}
    \includegraphics[width=\textwidth]{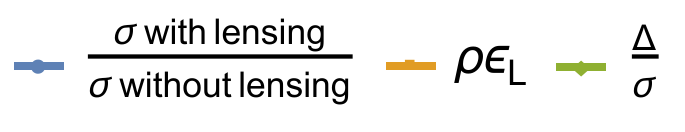}
  \end{subfigure}
  \end{center}
  \caption{Results for Euclid: we show the shifts in the best fit values in units of the errors from the Newtonian analysis (green line), the correlation with the lensing parameter (orange line) and the ratio between the errors with lensing to the one without lensing (blue line).}
    \label{fig:shiftEU}
\end{figure}

\begin{figure}[h!]
   \begin{subfigure}[h!]{0.345\textwidth}
    \includegraphics[width=\textwidth]{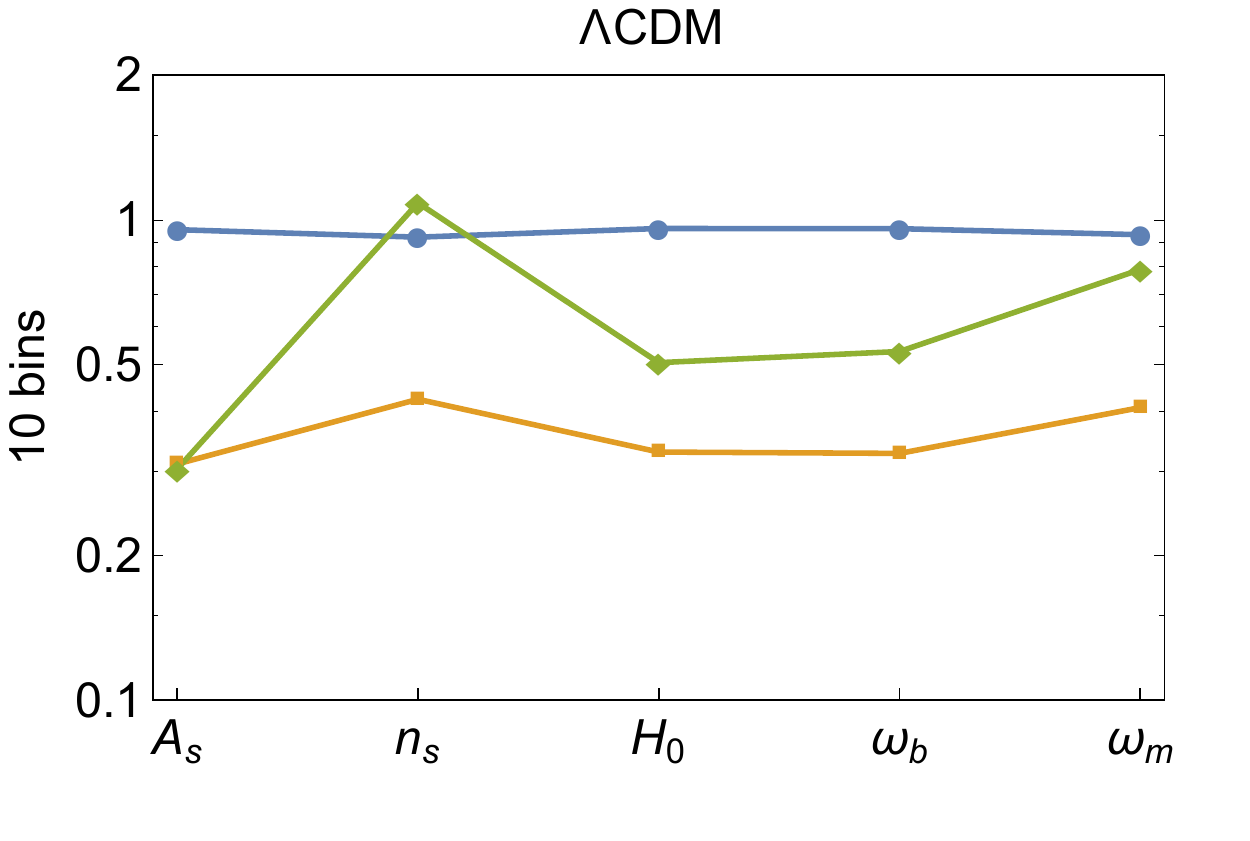}
  \end{subfigure}
   \hspace{-2.5cm}
  \begin{subfigure}[h!]{0.345\textwidth}
    \includegraphics[width=\textwidth]{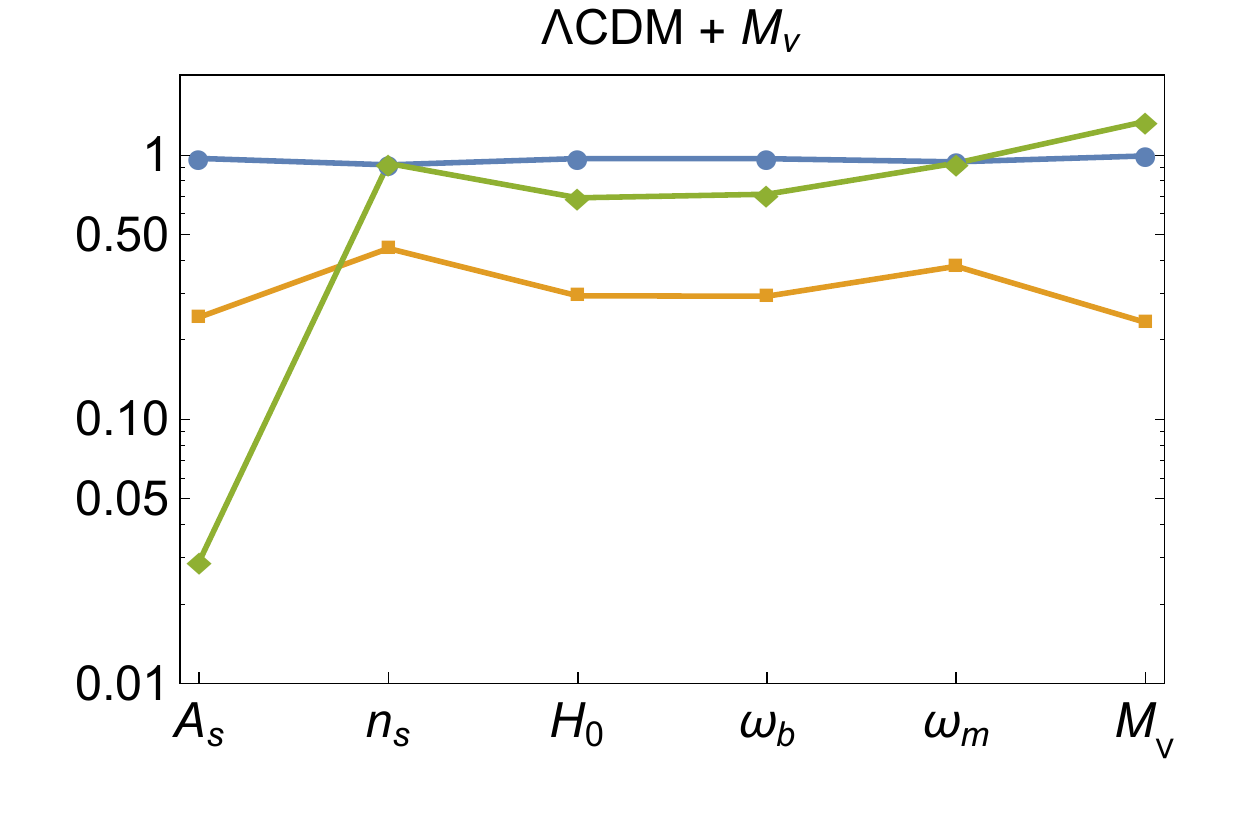}
  \end{subfigure}
   \hspace{-2.5cm}
  \begin{subfigure}[h!]{0.345\textwidth}
    \includegraphics[width=\textwidth]{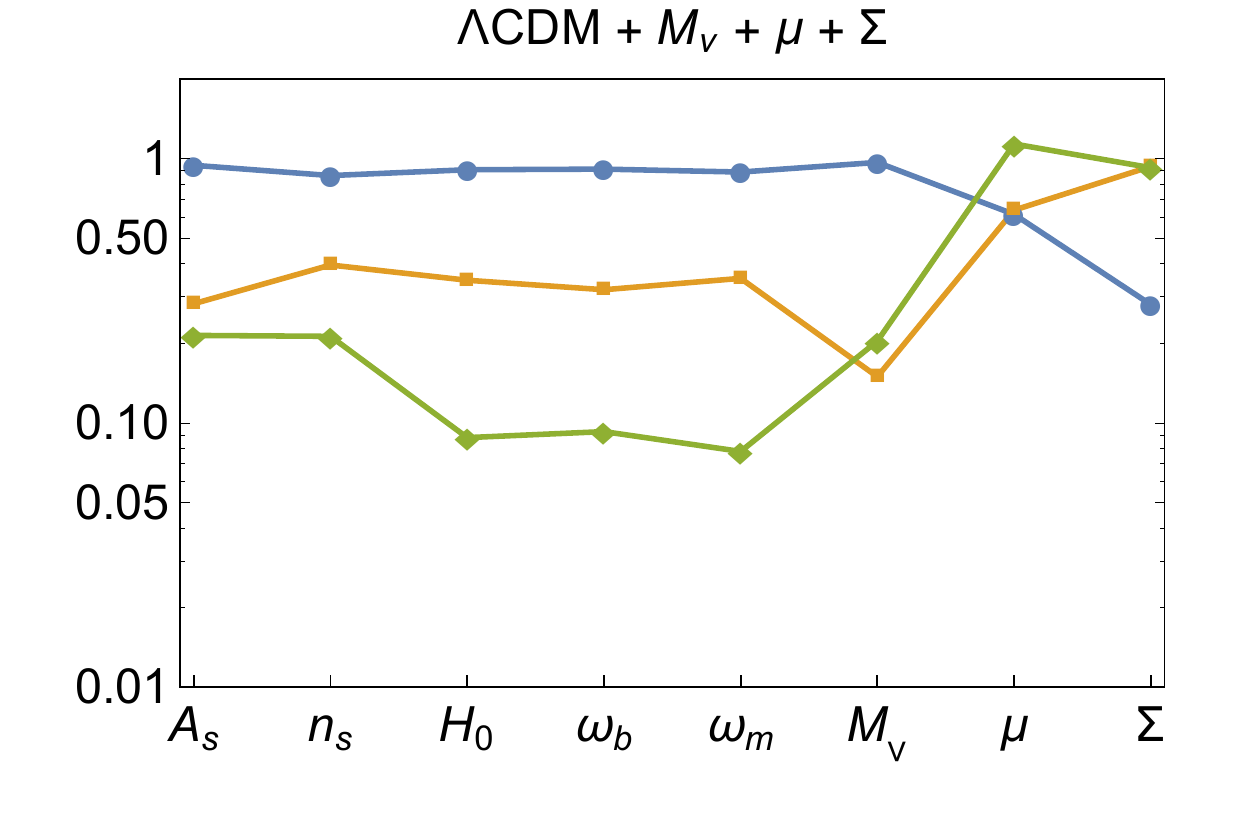}
  \end{subfigure}
    \begin{subfigure}[h!]{0.345\textwidth}
    \includegraphics[width=\textwidth]{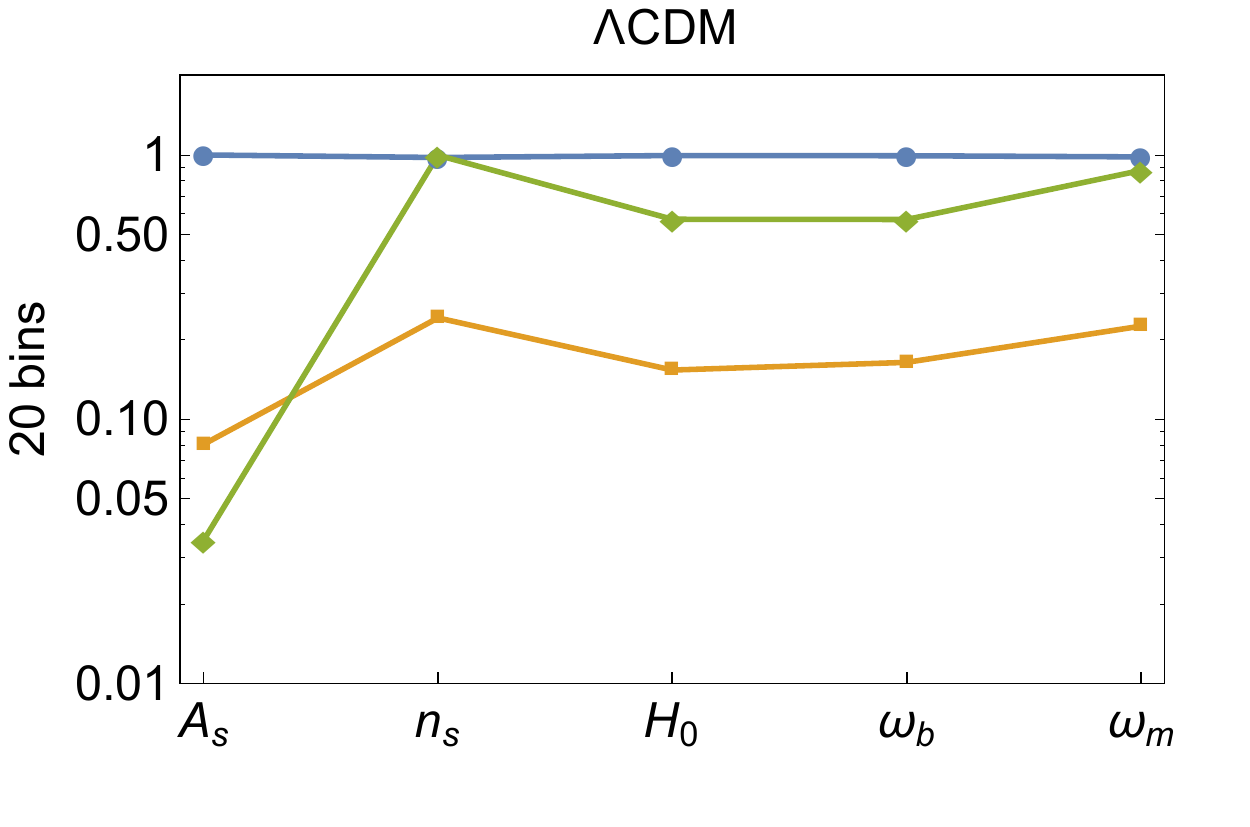}
  \end{subfigure}
   \hspace{-2.5cm}
  \begin{subfigure}[h!]{0.345\textwidth}
    \includegraphics[width=\textwidth]{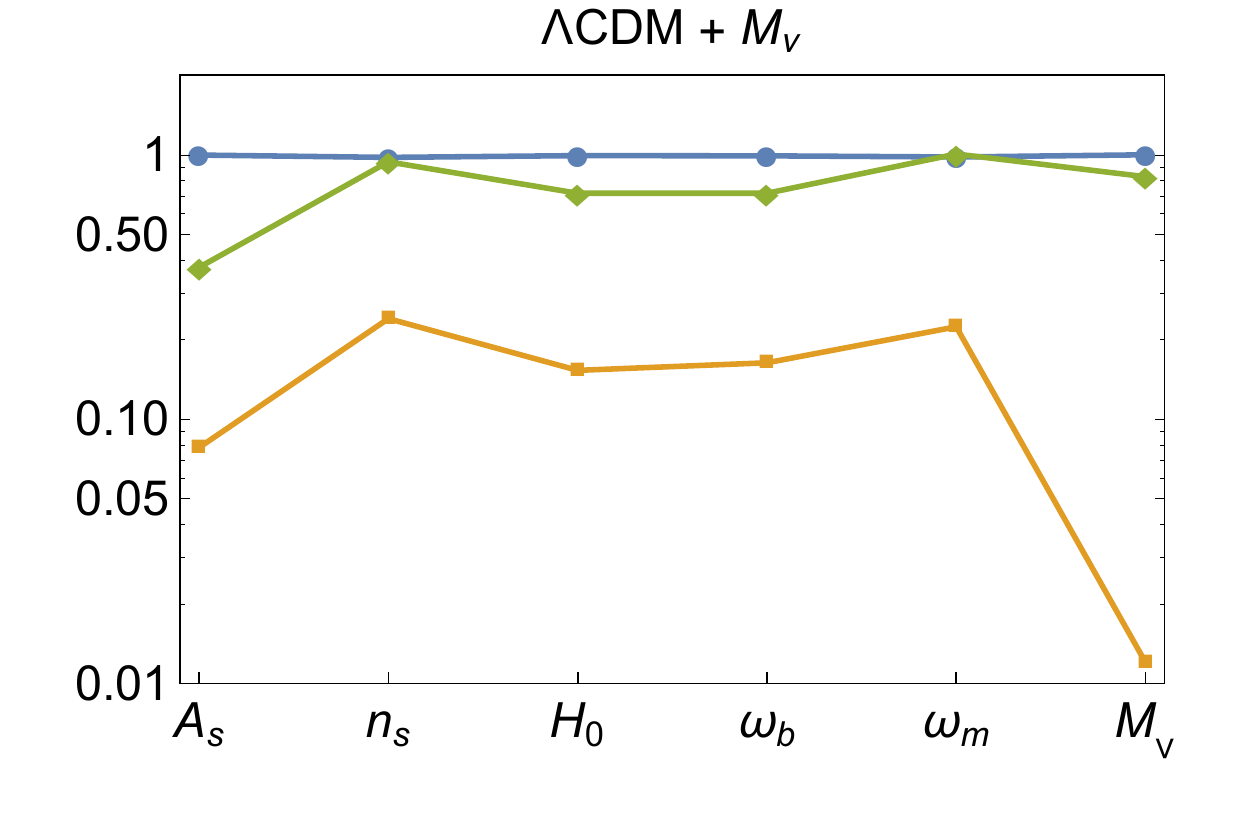}
  \end{subfigure}
   \hspace{-2.5cm}
  \begin{subfigure}[h!]{0.345\textwidth}
    \includegraphics[width=\textwidth]{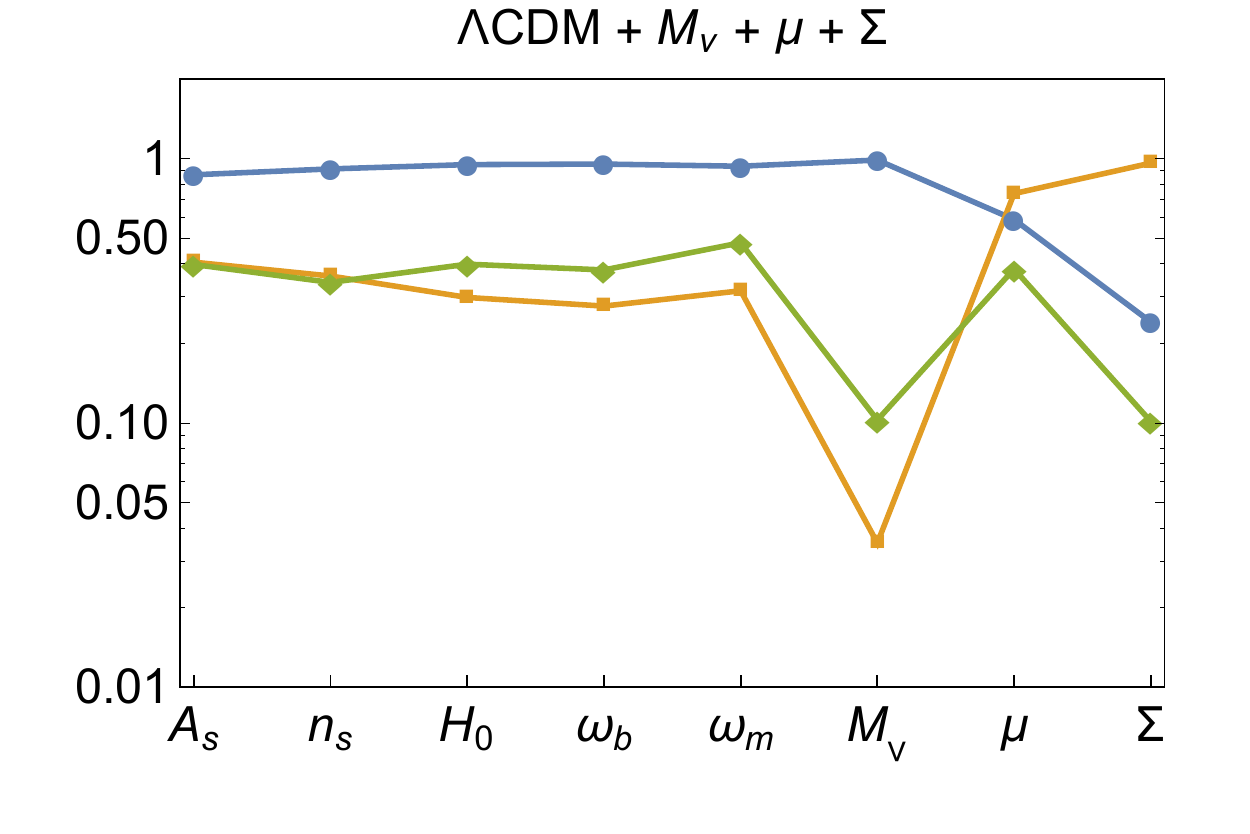}
  \end{subfigure}  
%%%%%%%%%%%%%%%%%%%%%%%%%%%%%%%%%%%%%%%%%%%%  
    \begin{subfigure}[h!]{0.345\textwidth}
    \includegraphics[width=\textwidth]{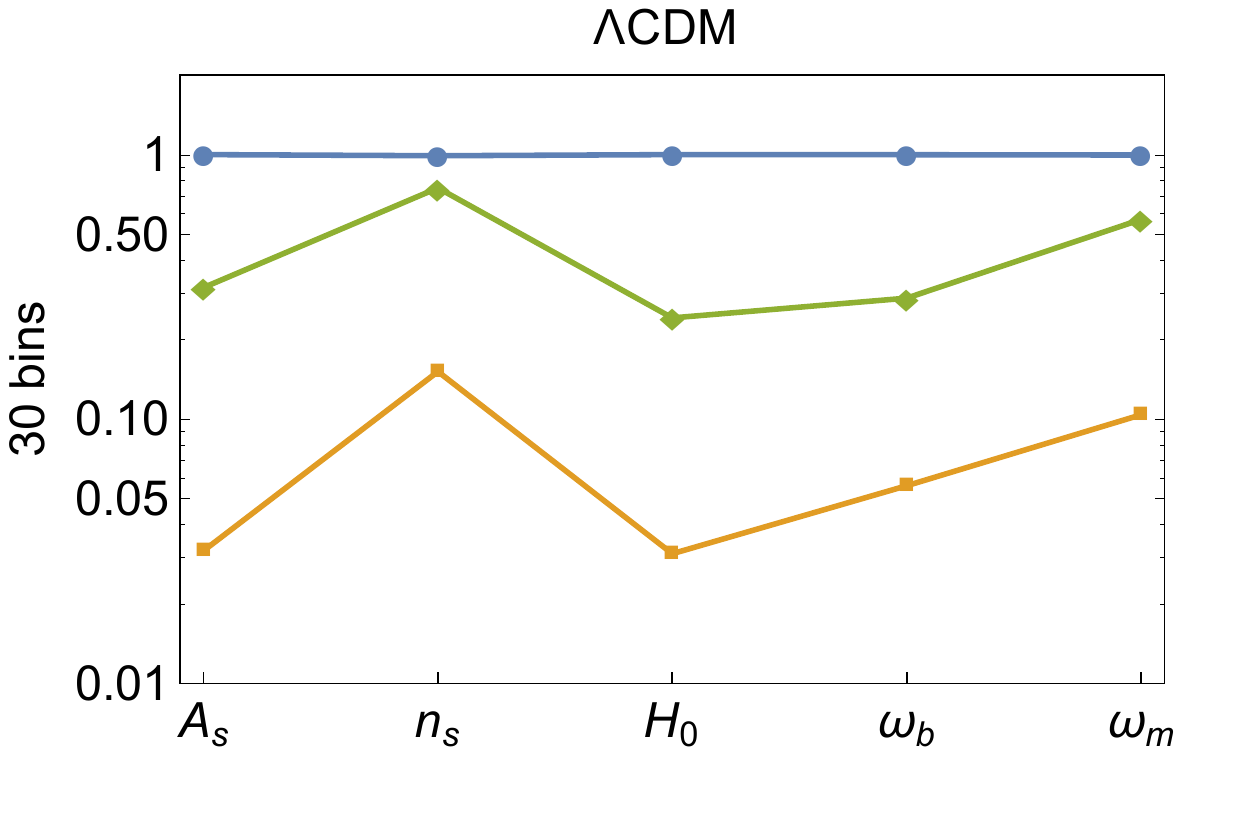}
  \end{subfigure}
   \hspace{-2.5cm}
  \begin{subfigure}[h!]{0.345\textwidth}
    \includegraphics[width=\textwidth]{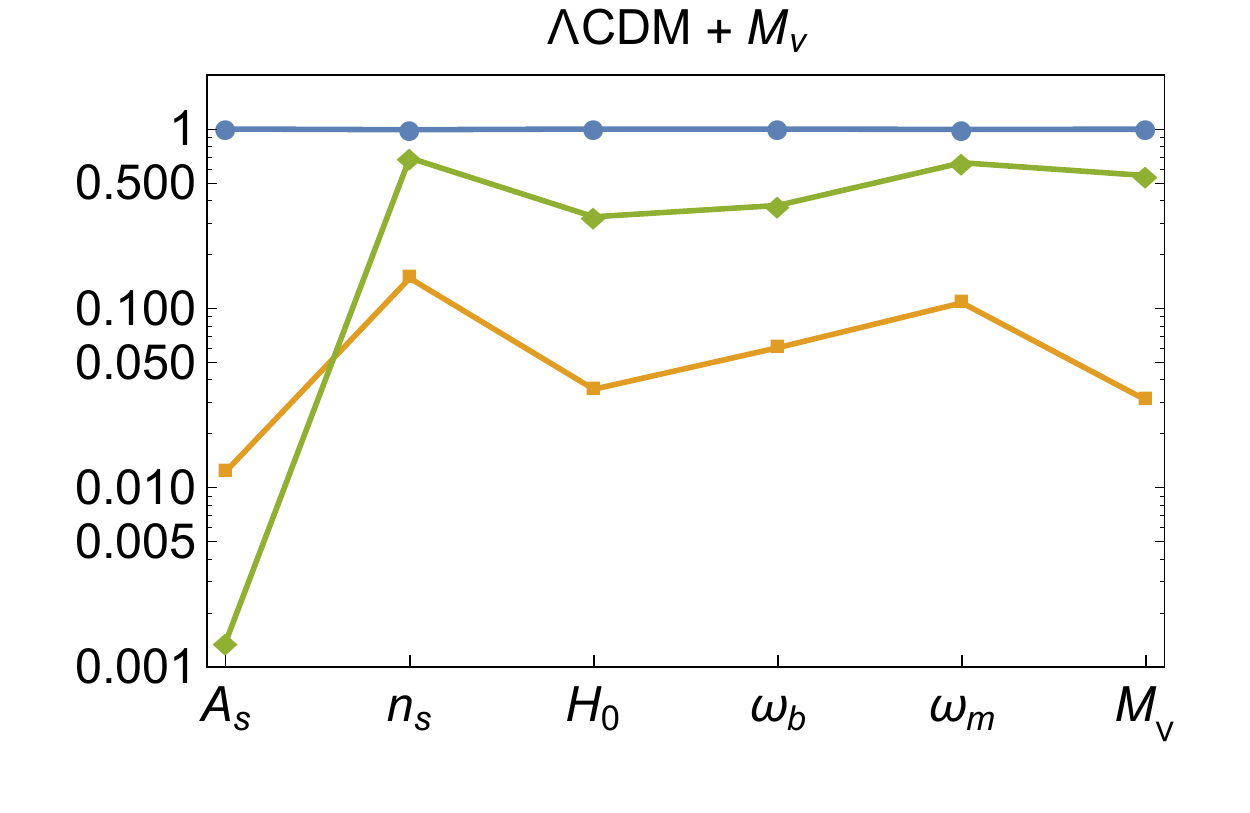}
  \end{subfigure}
   \hspace{-2.5cm}
  \begin{subfigure}[h!]{0.345\textwidth}
    \includegraphics[width=\textwidth]{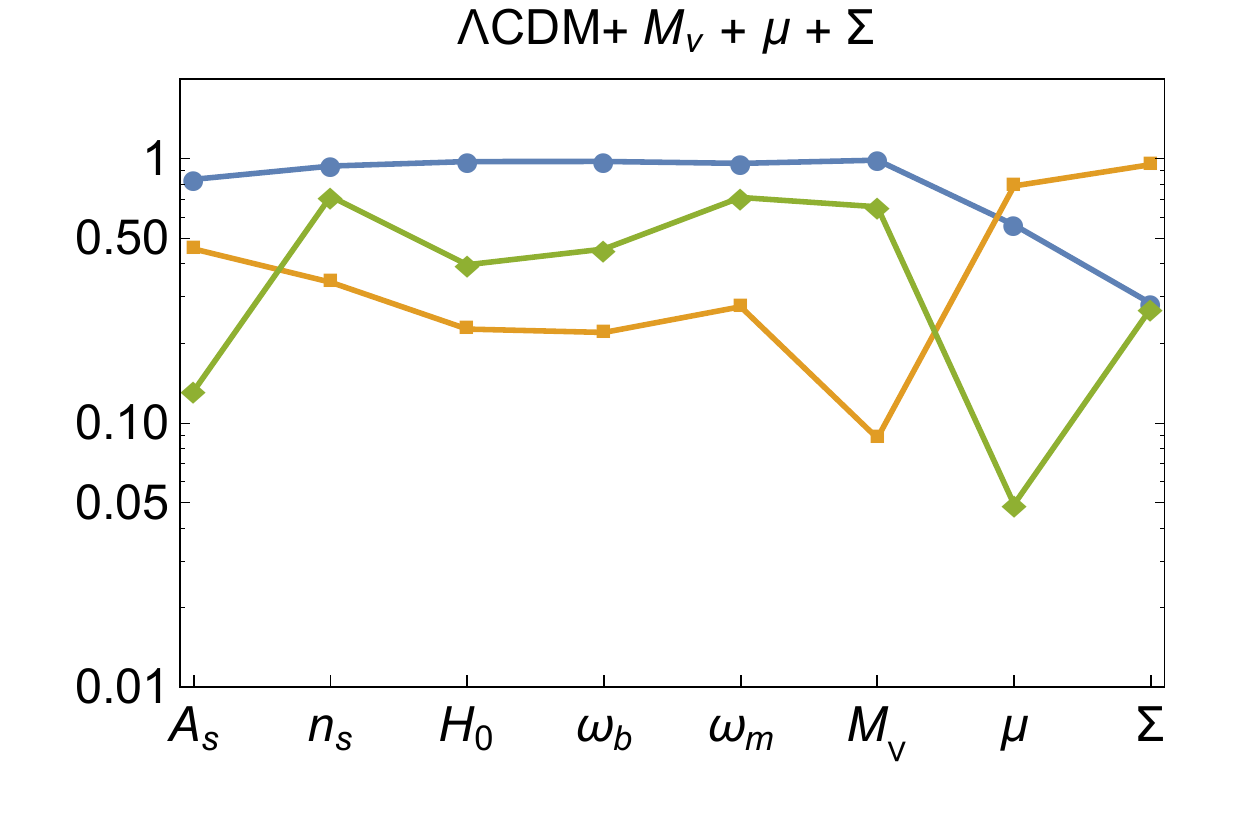}
  \end{subfigure}  
 %%%%%%%%%%%%%%%%%%%%%%%%%%%%%%%%%%%%%%%%%% 
   \begin{subfigure}[h!]{0.345\textwidth}
    \includegraphics[width=\textwidth]{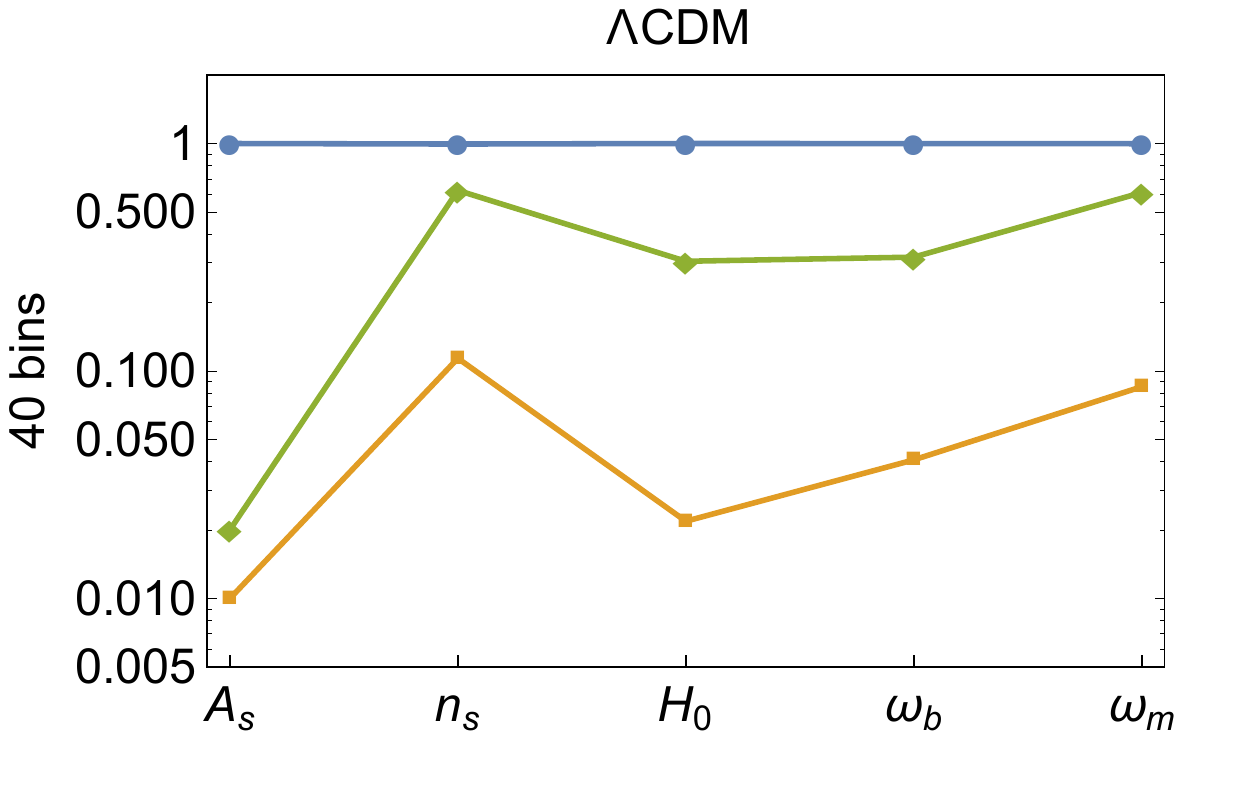}
  \end{subfigure}
   \hspace{-0.5cm}
  \begin{subfigure}[h!]{0.345\textwidth}
    \includegraphics[width=\textwidth]{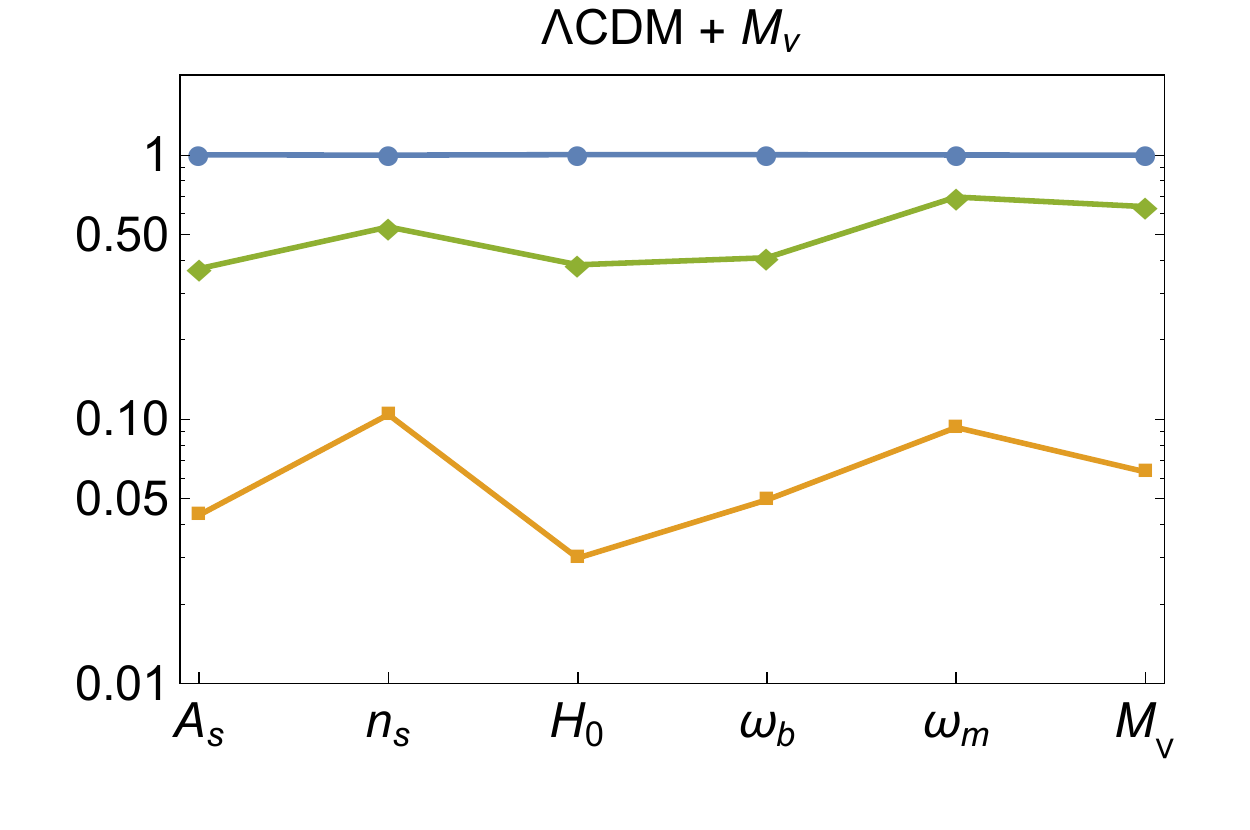}
  \end{subfigure}
   \hspace{-0.5cm}
  \begin{subfigure}[h!]{0.345\textwidth}
    \includegraphics[width=\textwidth]{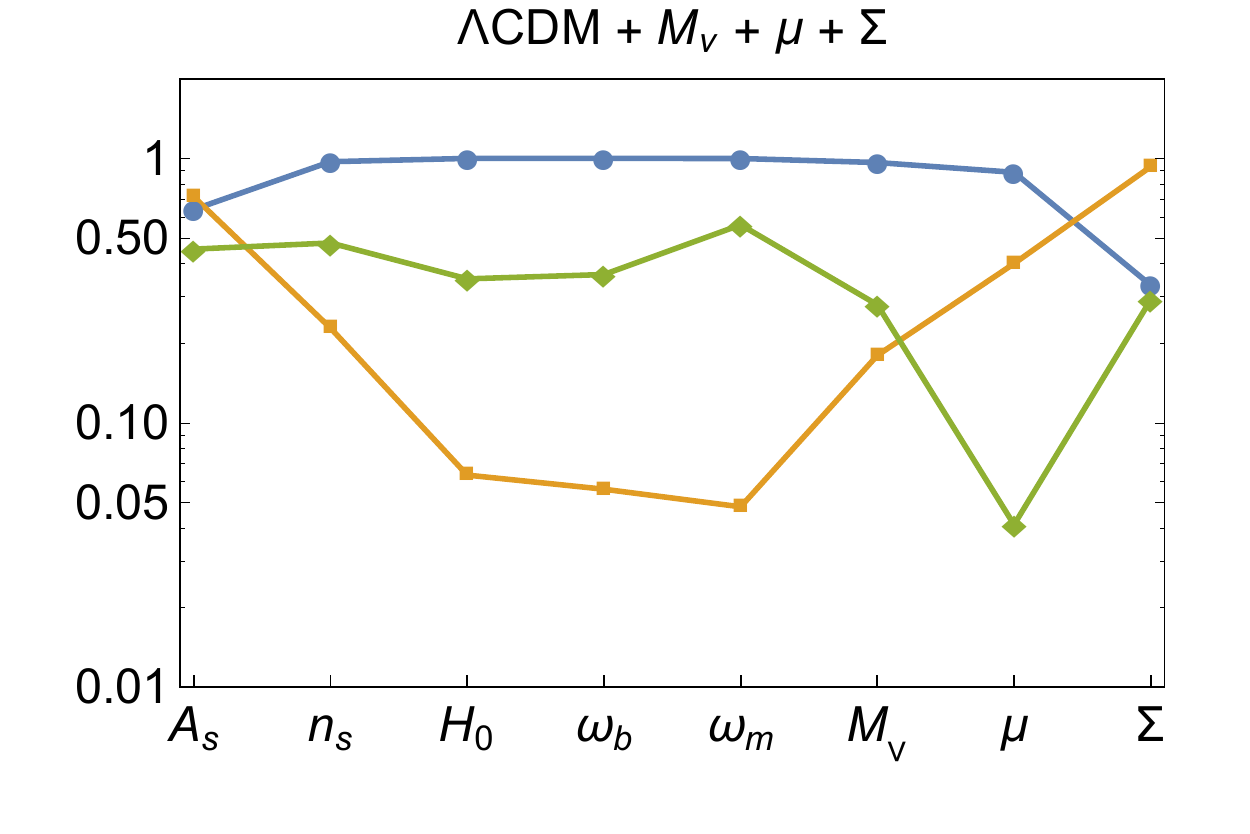}
  \end{subfigure}  
   \vspace{-0.4cm}
  \begin{center}
   \begin{subfigure}[h!]{0.3\textwidth}
    \includegraphics[width=\textwidth]{legendcorr.pdf}
  \end{subfigure}
  \end{center}
    \caption{Results for SKA: we show the shifts in the best-fit values in units of the errors from the Newtonian analysis (green line), the correlation with the lensing parameter (orange line) and the ratio between the errors with lensing to the one without lensing (blue line). Here we consider all the cross-correlations.}
 \label{fig:shiftSKA}
\end{figure}

In this section we present our results for the shift in the best-fit values of the $\Lambda$CDM and modified gravity parameters. We report in the same plots in Fig.~\ref{fig:shiftEU} and Fig.~\ref{fig:shiftSKA} for every parameter: its correlations with the lensing parameter, the ratio of the errors with lensing to the one obtained without and the results for the shift in the best-fit values. This is to highlight the correspondence that we expect between these three different quantities: the parameters whose correlation with the lensing parameter is higher are those whose estimation is more biased by neglecting lensing and also those whose constraints are more improved by including lensing. We express the shifts, given by Eq.~\eqref{shiftformula}, in units of the error we get from the standard Newtonian analysis. 
From the left and middle panels of Figs.~\ref{fig:shiftEU} and~\ref{fig:shiftSKA}, it is evident the one-to-one correspondence between improvement on the errors, correlation with lensing and shifts, both Euclid and SKA, for the $\Lambda$CDM model and for the extension including massive neutrinos. When we add modified gravity the values of the shifts is not completely determined by the correlation with the lensing parameter because it is influenced also by the correlations between the parameters of the model itself, which are altered in modified gravity. 

For Euclid, and in general for photometric surveys, we find shifts larger than $1\sigma$, and even more for the modified gravity parameters. We again explain this behaviour considering that photometric surveys have a much better angular resolution than radial, therefore they are more sensitive to the lensing effects and by neglecting its contribution in the analysis will affect the cosmological parameters process stronger.
Note that for the $\Lambda$CDM and for the $\Lambda {\rm CDM} + M_\nu$ models the shifts on the cosmological parameters are all around $1\sigma$, reaching $2\sigma$ for the spectral index $n_s$ and neutrino mass $M_\nu$, whereas they drop to smaller values when adding modified gravity, see Tables~\ref{tableLCDMn} and~\ref{tableLCDMnMG} in Appendix~\ref{sec:table}.

For SKA, for the $\Lambda$CDM and for the extension with neutrinos the shifts on all the cosmological parameters are below  $1\sigma$ and sistematically lower than the ones for Euclid, as expected, see left and middle panels of Fig.~\ref{fig:shiftSKA} compared with those of Fig.~\ref{fig:shiftEU}.
When including modified gravity, the bias on the cosmological parameters and neutrino mass are lower than in the $\Lambda$CDM and the $\Lambda$CDM + $M_\nu$ models and we find a bias of more than $1\sigma$ on $\mu$ and $\Sigma$ only for the 10 bins configuration.
In this respect, it is worth remarking the dependence of the shifts on the modified gravity parameters on the number of bins: for 10 bins the lensing signal is such that the shifts on $\mu$ and $\Sigma$ are more than $1\sigma$. But by increasing the number of redshift bins the shifts on $\mu$ and $\Sigma$ reflect the relative behaviour of the constraining power of lensing and redshift-space distortions that we have already remarked in section~\ref{sec:Rconstraints}. In particular we find that when we approach a number of bins where the angular analysis is competitive with respect to the full 3-dimensional analysis, i.e. when the width of the bin is comparable to the non-linear scale~\cite{Asorey:2012rd,DiDio:2013sea}, the shift are below 1$\sigma$. We may conclude that SKA is not sensitive to any bias induced by neglecting lensing magnification. Nevertheless we want to stress that the shift on the parameter $\Sigma$ if expressed in units of the error obtained including lensing is about 1$\sigma$, whereas the shifts on the other cosmological parameters remain unchanged.

\begin{figure}[t]
%%%%%%%%%%%%%%%%%%%%%%%%%%%%%%%%%%%%%%%%%%%%%%  
  \begin{subfigure}[h!]{0.4\textwidth}
    \includegraphics[width=\textwidth]{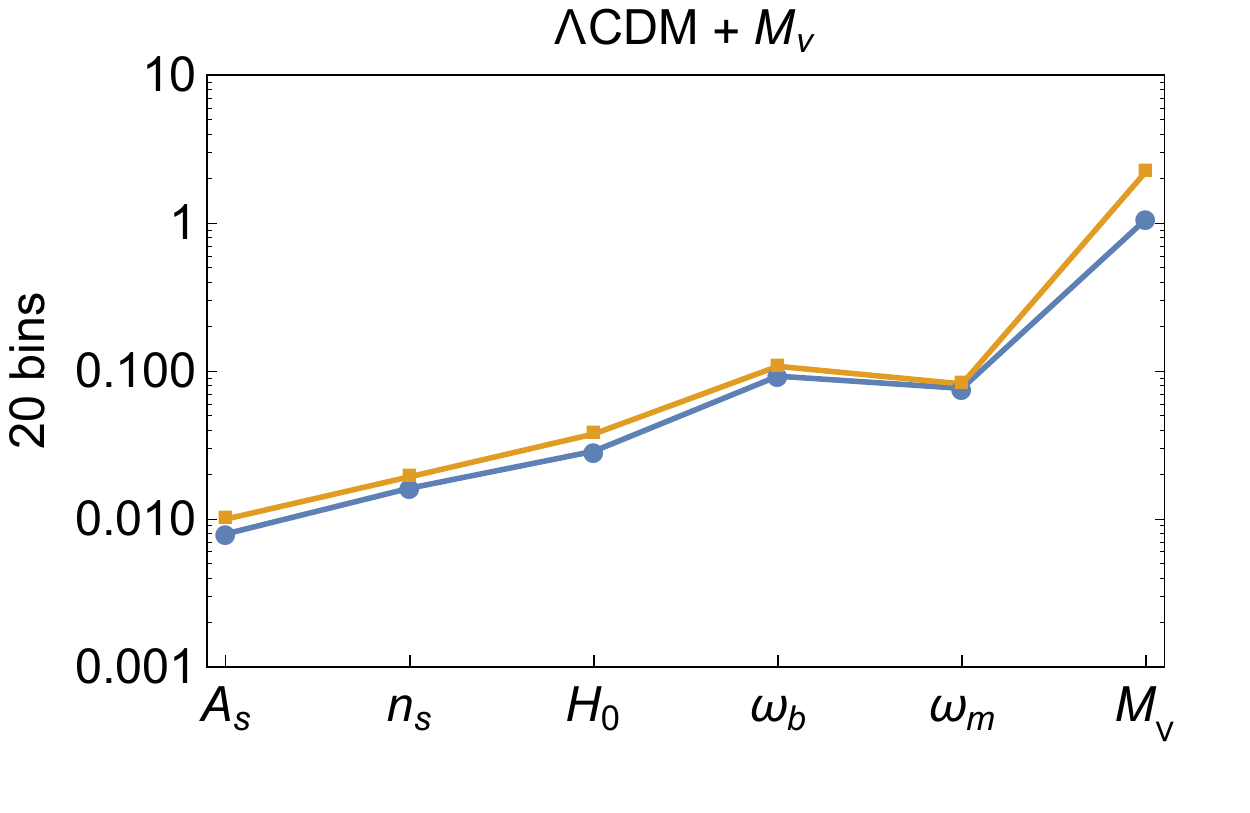}
  \end{subfigure}
  \hspace*{-0.5cm}
  \begin{subfigure}[h!]{0.4\textwidth}
    \includegraphics[width=\textwidth]{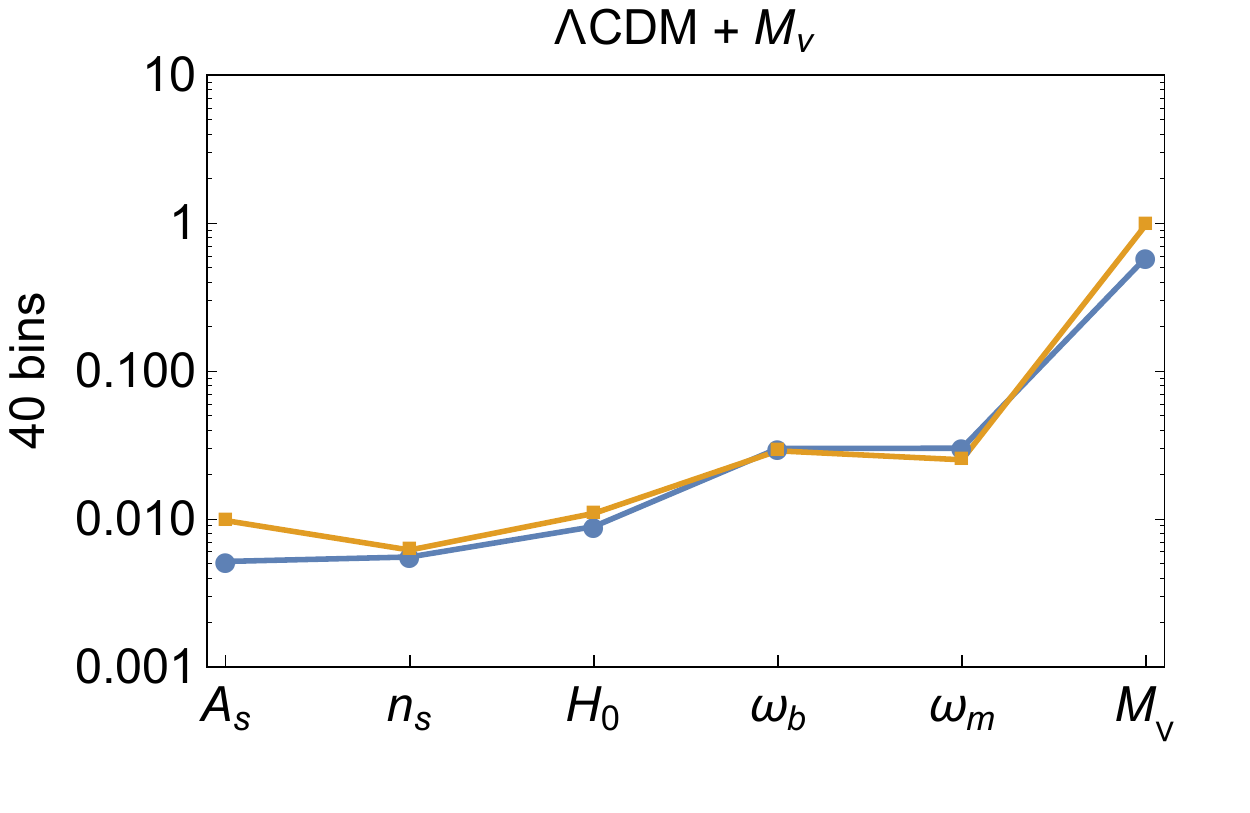}
  \end{subfigure} 
  \hspace*{-0.5cm}
   \begin{subfigure}[h!]{0.15\textwidth}
    \includegraphics[width=\textwidth]{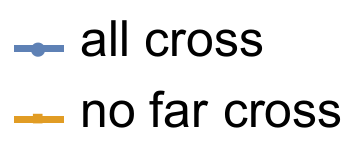}
  \end{subfigure} 
  %----------------------------------------------------------------------------------------------------------
    \begin{subfigure}[h!]{0.4\textwidth}
    \includegraphics[width=\textwidth]{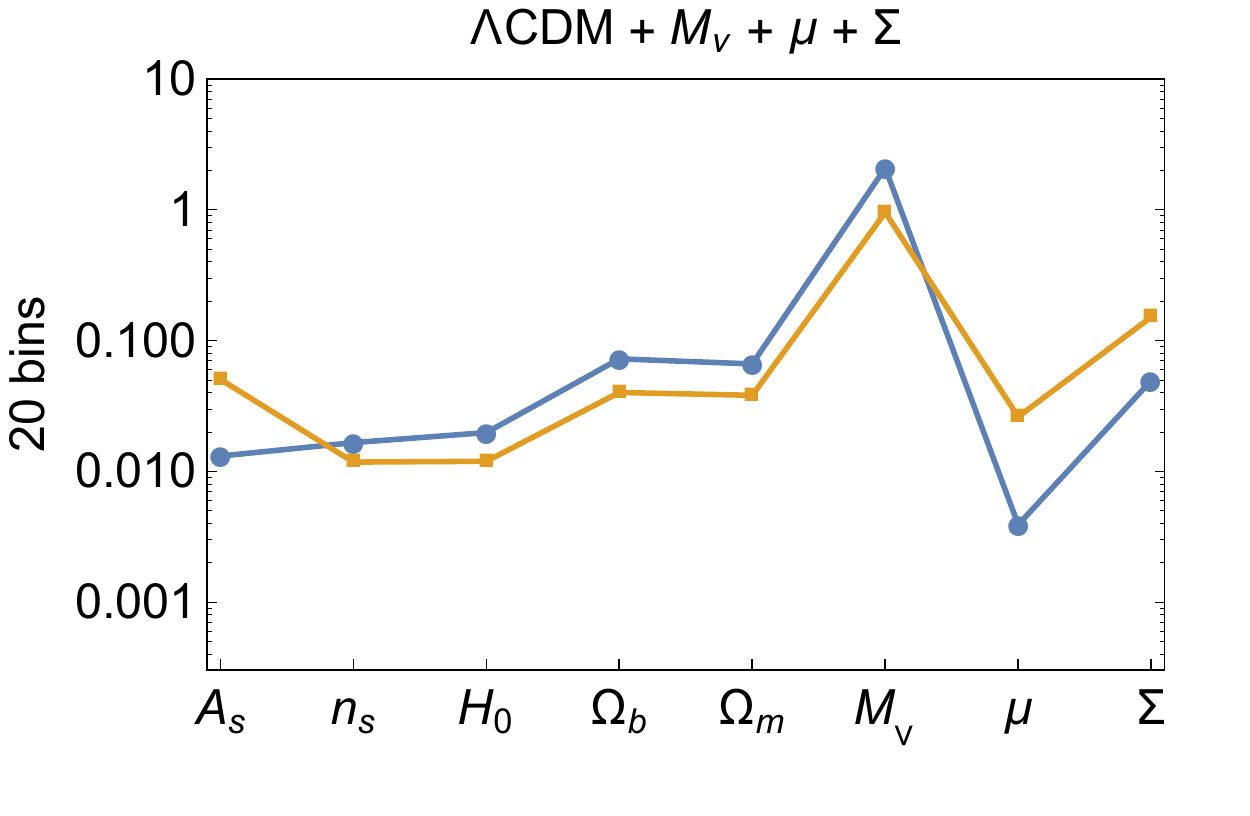}
  \end{subfigure}
  \hspace*{0.1cm}
  \begin{subfigure}[h!]{0.4\textwidth}
    \includegraphics[width=\textwidth]{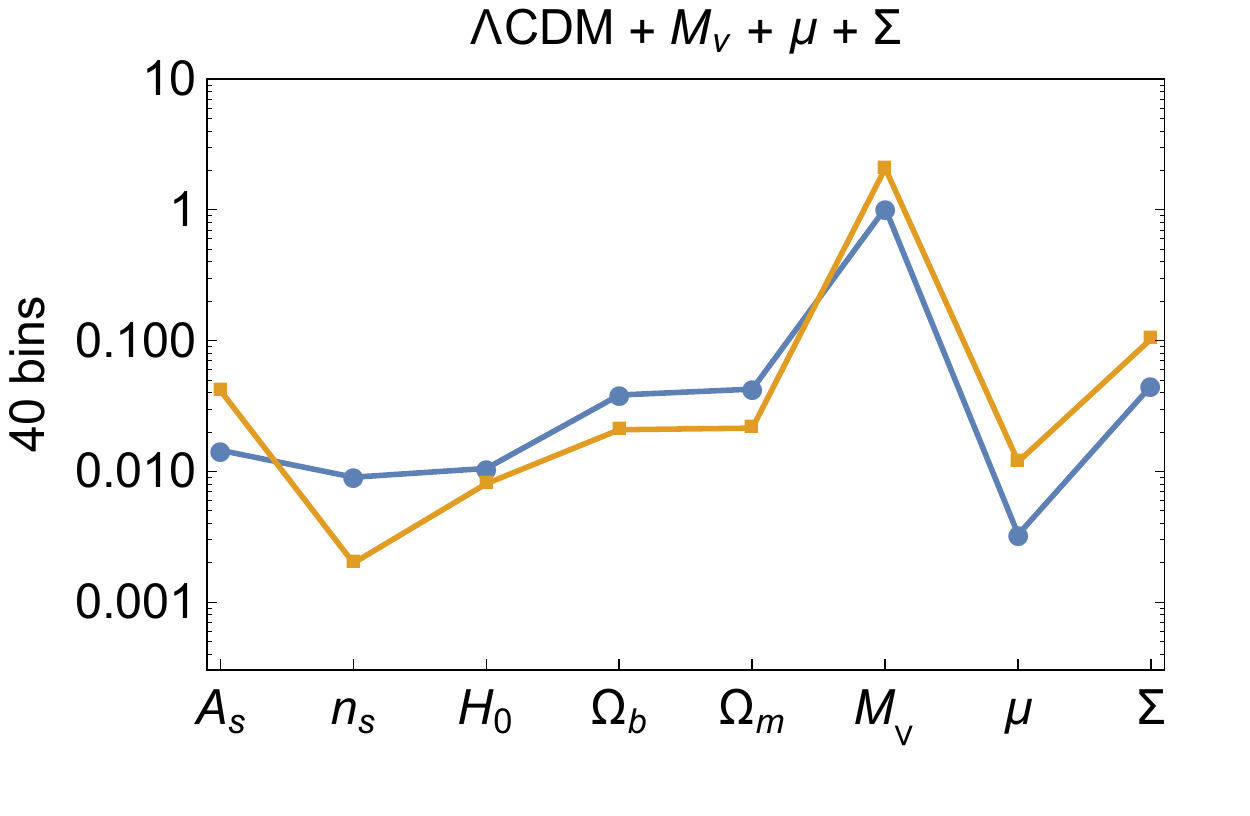}
  \end{subfigure}  
  \hspace*{0.1cm}
   \begin{subfigure}[h!]{0.15\textwidth}
    \includegraphics[width=\textwidth]{legendshift.pdf}
  \end{subfigure}  
    \caption{Shifts in the best-fit values of the parameters for the two extensions of the $\Lambda$CDM model for SKA. We compare the results obtained by including all the cross-correlations (blue line) and those obtained by neglecting the far-bin cross-correlations (orange line). The shifts are given in units of the fiducial values for both the configurations.}
    \label{fig:crfshiftSKA}
    %%%%%%%%%%%%%%%%%%%%%%%%%%%%%%%%%%%%%%%%%%%%%%%%%%%%%%%%%
\end{figure}

As part of our analysis for SKA we also compare the shifts that we get by considering all the cross-correlations with those obtained by setting to zero those between far redshift bins. From Fig.~\ref{fig:crfshiftSKA} we see that we basically obtain the same results with both configurations and for all the parameters, including $\mu$ and $\Sigma$. This means that the lensing contribution from far-bin cross-correlations is sub-dominant for the estimation of the bias of the best-fit value, despite the lensing convergence dominates the signal of the angular power spectra of these cross-correlations.

\begin{figure}[h!]
%%%%%%%%%%%%%%%%%%%%%%%%%%%%%%%%%%%%%%%%%%%%%% %%%%%%%%%%%%%%%%%% 
  \begin{subfigure}[h!]{0.42\textwidth}
    \includegraphics[width=\textwidth]{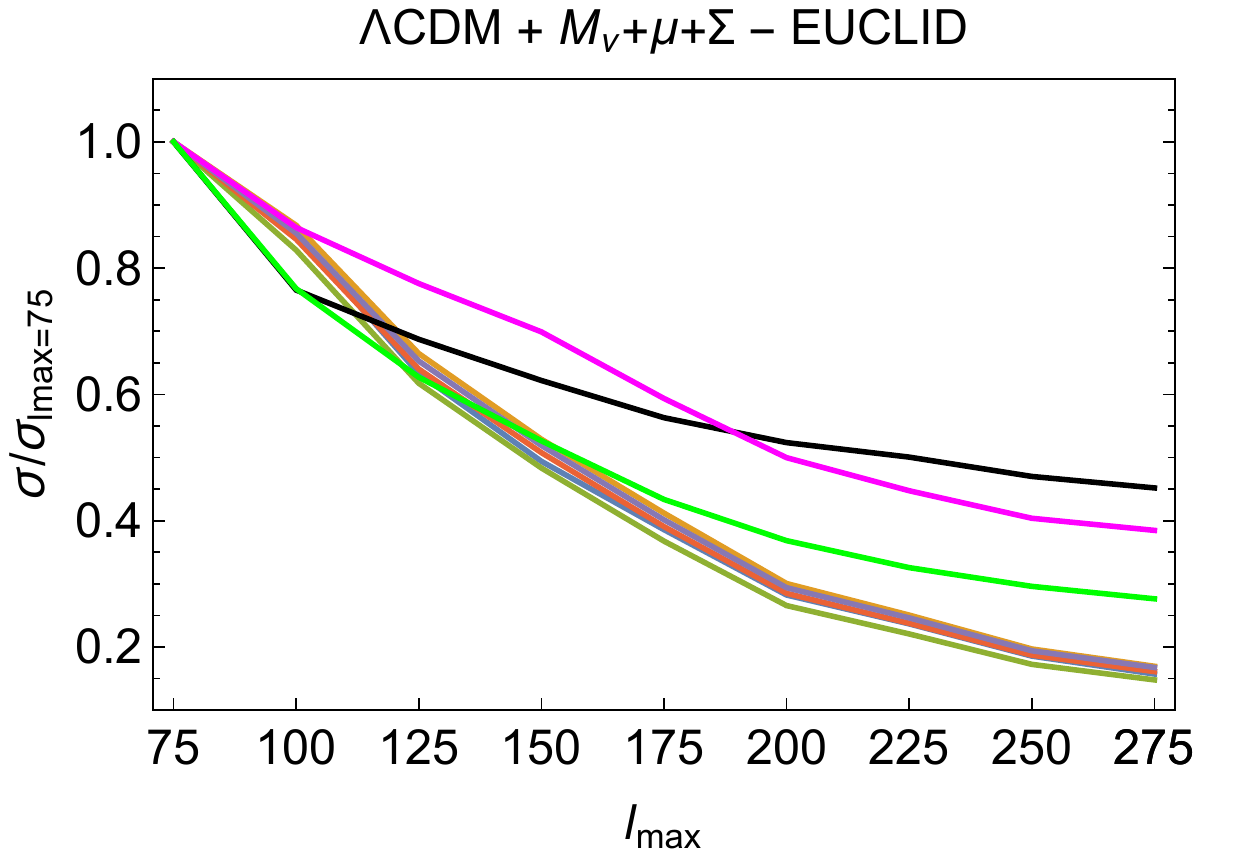}
  \end{subfigure}
%  \hspace*{-0.5cm}
  \begin{subfigure}[h!]{0.42\textwidth}
    \includegraphics[width=\textwidth]{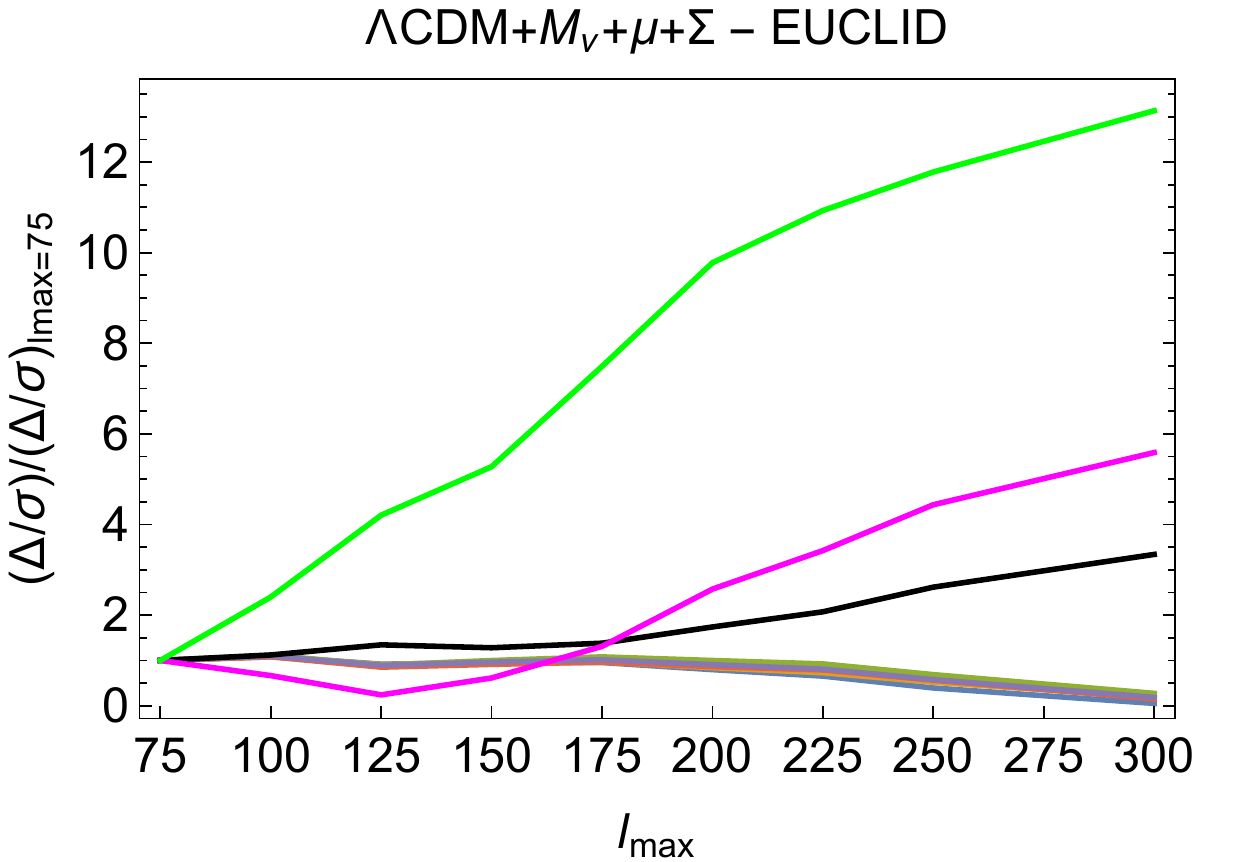}
  \end{subfigure} 
 % \hspace*{-0.5cm}
   \begin{subfigure}[h!]{0.08\textwidth}
    \includegraphics[width=\textwidth]{mblegendsigma.pdf}
  \end{subfigure} 
  \caption{Results for Euclid in the 10 bins configuration. Left panel: errors for the cosmological and modified gravity parameters for different values of the maximum multipole. Right panel: shift in the best-fit values of the cosmological and modified gravity parameters in units of the standard error for different values of the maximum multipole. Both are normalized to $\ell_{\rm {max}}=75$. Here the errors are calculated with the Newtonian theoretical power spectra and lensing in the covariance.}
  \label{fig:EUlmax}
  %%%%%%%%%%%%%%%%%%%%%%%%%%%%%%%%%%%%%%%%%%%%%%%%%%%%%%%%%%%%%%%
  \end{figure}

Let us finally comment on the dependence of the bias in parameter estimation on the choices for the maximum multipole. In this case we consider Euclid, which exhibits the most significant shifts. In Fig.~\ref{fig:EUlmax} we show the dependence of the errors (left panel) and the dependence of the shifts (in units of the errors, right panel) on the maximum multipole $\ell_{\rm max}$. For both the quantities it is clear the separation in two groups: neutrino mass and modified gravity parameters at one side and all the parameters of the standard $\Lambda$CDM model on the other side. Given that if we include more multipoles the errors decreases for each parameter, from the left panel we see that for neutrino mass, $\mu$ and $\Sigma$ the information coming from including larger multipoles is less significant, meaning that they are more sensitive to larger scales. 
We also remark that on large scale there is a more relevant contribution due to the pure lensing term, while on smaller scales the lensing signal is dominated by its correlation with the standard Newtonian terms.
This behaviour is mirrored in their shifts: the major contribution comes from larger scale and they do not change significantly as $\ell_{\rm max}$ increases. Therefore, once the shifts are expressed in terms of the errors, the information gained from smaller scales does not compensate the bias from larger scales and the result is that the values of $\Delta/\sigma$ increases, as show in the right panel of Fig.~\ref{fig:EUlmax}.

%%%%%%%%%%%%%%%%%%%%%%%%%%%%%%%%%%%%%%%%%%%%%%%%%%%%%%%%%%%%%%%%%%%%%%%
%%%%%%%%%%%%%%%%%%%%%%%%%%%%%%%%%%%%%%%%%%%%%%%%%%%%%%%%%%%%%%%%%%%%%%%
\section{Conclusions}
\label{sec:conclusions}
%%%%%%%%%%%%%%%%%%%%%%%%%%%%%%%%%%%%%%%%%%%%%%%%%%%%%%%%%%%%%%%%%%%%%%%
%%%%%%%%%%%%%%%%%%%%%%%%%%%%%%%%%%%%%%%%%%%%%%%%%%%%%%%%%%%%%%%%%%%%%%%
In this paper we have investigated the impact of neglecting lensing magnification in future galaxy clustering surveys, motivated by the deeper redshift range probed by them. 
We consider three cosmological models, namely the standard $\Lambda$CDM model and two extensions: the first including massive neutrinos and the second including deviations to General Relativity. 
We focus both on photometric and spectroscopic redshift surveys, showing their different behaviour regard lensing magnification. 

In particular, photometric surveys due to the poor redshift resolution measure mainly transversal modes which are more affected by lensing magnification. This leads to a larger bias induced by neglecting lensing magnification, in particular for the terms beyond $\Lambda$CDM. We show that the cosmological parameters which exhibit a larger bias are the ones more correlated with lensing magnification. This is clear for the parameters ($\mu$ and $\Sigma$) which parametrise deviations from General Relativity, where a crucial information is carried by the lensing potential.

To fully consider the impact on spectroscopic surveys we would need a number of redshift bins such that the width of them is comparable with the non-linear scale. While this imposes strong numerical limitations, we show the behaviour by increasing the number of redshift bins, hence by approaching a spectroscopic resolution. By doing that we including more and more radial modes which are unaffected by lensing (being a pure transversal effect) and therefore the relative information carried by lensing potential is less relevant. Interestingly this does not apply completely to the $\Sigma$ parameter, whose information is mainly carried by the lensing potential. 
As well known redshift space distortion measures the growth of structures, and it can be used to constrain the underling theory of gravity. In our analysis we show that when we start approaching a good enough redshift resolution and we begin to probe (small) scale radial modes, the constraining power on the $\mu$ parameter is dominated by redshift space distortion compared to lensing magnification.
This different behaviour between $\Sigma$ and $\mu$ reflects the modification of eqs.~\eqref{mu} and~\eqref{sigma}. 

We have also shown that, while lensing magnification dominates the cross-correlation between far away redshift bins, the information encoded in such correlations is very subdominant with respect to closer redshift correlations and by including or not far cross-correlations does not change the results.

We have therefore shown that it is important to include lensing magnification in data analysis, in order to use next generation of clustering surveys to test for deviations from General Relativity. In particular, our results show that the estimation of the modified gravity parameters is biased for photometric surveys whereas the shift in the best-fit values stay below $1\sigma$ for spectroscopic surveys.
\vspace{1cm}
\\
{\bf Note added} \\
Some similar issues on the same topic are addressed in \citep{Lorenz:2017iez} which appeared while this work was in the final stage of preparation. Wherever possible we verified that our results agree with theirs.
We emphasise that our results address the issues for spectroscopic surveys as well.

\acknowledgments

The authors acknowledge financial support by the INFN-INDARK grant IS PD51. ED is supported by the cosmoIGM starting grant and from the Swiss National Science Foundation (No.~171494).
We thank Francesco Montanari and Alvise Raccanelli for helpful suggestions and David Alonso, Stefano Camera and Miguel Zumalac\' arregui for useful discussions. %

%%%%%%%%%%%%%%%%%%%%%%%%%%%%%%%%%%%%%%%%%%%%%%%%%%%%%%%%%%%%%%%%%%%%%%%%%%%%%%%%%%%%%%%%%%%
\appendix
%%%%%%%%%%%%%%%%%%%%%%%%%%%%%%%%%%%%%%%%%%%%%%%%%%%%%%%%%%%%%%%%%%%%%%%%%%%%%%%%%%%%%%%%%%%

\section{Summary of the results}
\label{sec:table}
We show here the values that we obtain for the correlation with the lensing parameter, the ratio of the errors for the cosmological and modified gravity parameters - errors with lensing to errors without lensing - and the shift in their best-fit value expressed in unit of the sigma obtained from a Newtonian analysis. We report our results for the 10 bins configuration for Euclid and for the 40 bins for SKA.

\begin{table}[h!]%[tbp]
\centering
\begin{tabular}{|c|ccc|ccc|}
\hline
\multicolumn{1}{|c}{}              & 
\multicolumn{3}{c|}{EUCLID - 10 bins}              & 
  \multicolumn{3}{|c|}{SKA - 40 bins}             \\
  \hline
 \multicolumn{1}{|c}{}                    & 
 \multicolumn{1}{c}{$\rho_{\epsilon_\text{L}}$}&
 \multicolumn{1}{c}{$\sigma_\text{lens}/\sigma_\text{w/o lens}$} &
 \multicolumn{1}{c|}{$\Delta/\sigma$}   &
  \multicolumn{1}{c}{$\rho_{\epsilon_\text{L}}$}&
 \multicolumn{1}{c}{$\sigma_\text{lens}/\sigma_\text{w/o lens}$} &
 \multicolumn{1}{c|}{$\Delta/\sigma$}                    \\
\hline  
$A_s$             &  0.07617 & 1.000&  -0.6094 & 0.005150& 0.9990&0.3705 \\
$n_s$             &  0.44019 & 0.9118& -1.753   & -0.005522&0.9966&-0.5332 \\
$H_0$            &  -0.2958 & 0.9632&   1.206 &  0.008866&1.001& 0.3834\\
$\omega_b$  &  -0.2991  & 0.9619&   1.183 & 0.02988 &1.000&0.4077 \\
$\omega_m$ &  -0.3711  & 0.9389&   1.480  & 0.03003 &0.9977& 0.6918\\
$M_{\nu}$      &   0.3923  & 0.9440&  -1.996 &0.5854 &0.9965&  0.6359\\            
\hline
\end{tabular}
\caption{Summary of results for the extended $\Lambda$CDM model ($\Lambda$CDM $+ M_{\nu}$). Here we report the results for SKA obtained by considering all the cross-correlations.
}
\label{tableLCDMn}
\end{table}

\begin{table}[h!]%[tbp]
\centering
\begin{tabular}{|c|ccc|ccc|}
\hline
\multicolumn{1}{|c}{}              & 
\multicolumn{3}{c|}{EUCLID - 10 bins}              & 
  \multicolumn{3}{|c|}{SKA - 40 bins}             \\
  \hline
 \multicolumn{1}{|c}{}                    & 
 \multicolumn{1}{c}{$\rho_{\epsilon_\text{L}}$}&
 \multicolumn{1}{c}{$\sigma_\text{lens}/\sigma_\text{w/o lens}$} &
 \multicolumn{1}{c|}{$\Delta/\sigma$}   &
  \multicolumn{1}{c}{$\rho_{\epsilon_\text{L}}$}&
 \multicolumn{1}{c}{$\sigma_\text{lens}/\sigma_\text{w/o lens}$} &
 \multicolumn{1}{c|}{$\Delta/\sigma$}                    \\
\hline  
$A_s$            &   -0.1445       &   0.9806       &      -0.1079          & -0.7145     &    0.6403    &   0.4535 \\
$n_s$             &    0.3504      &   0.9033        &     -0.1467           & 0.2279      &    0.9709    &  -0.4784 \\
$H_0$            &   -0.2219      &   0.9647        &      0.2922           & -0.06334    &    0.9990    &   0.3500\\
$\omega_b$  &  -0.2030      &    0.9661        &      0.1822           & -0.05601   &   0.9990    &   0.3634 \\
$\omega_m$ &  -0.2420      &   0.9511         &      0.2259           &  -0.04806  &    0.9983     &   0.5603 \\
$M_{\nu}$      &   0.1209       &    0.9838       &      -2.112             & -0.1797    &     0.9637     &   0.2793 \\
$\mu$            &   -0.3055      &    0.8646          &     -1.996             & -0.3998    &     0.8852   &   0.04112 \\
$\Sigma$     &   -0.7867     &    0.5042         &     -1.878             &-0.9299     &      0.3312   &   0.2919 \\              
\hline
\end{tabular}
\caption{Summary of results for the extended $\Lambda$CDM model ($\Lambda$CDM $+ M_{\nu} + \mu + \Sigma$). Here we report the results for SKA obtained by considering all the cross-correlations.
}
\label{tableLCDMnMG}
\end{table}

\bibliographystyle{JHEP}
\bibliography{mybib}

\end{document}